\documentclass[twocolumn,preprint]{aastex631}

\usepackage{amsmath}

\usepackage{multirow}

\begin{document}

\title{Red Supergiant problem viewed from the nebular phase spectroscopy of type II supernovae}

\author[0000-0002-1161-9592]{Qiliang Fang}
\affiliation{National Astronomical Observatory of Japan, National Institutes of Natural Sciences, 2-21-1 Osawa, Mitaka, Tokyo 181-8588, Japan}
\author[0000-0003-1169-1954]{Takashi J. Moriya}
\affiliation{National Astronomical Observatory of Japan, National Institutes of Natural Sciences, 2-21-1 Osawa, Mitaka, Tokyo 181-8588, Japan}
\affiliation{Graduate Institute for Advanced Studies, SOKENDAI, 2-21-1 Osawa, Mitaka, Tokyo 181-8588, Japan}
\affiliation{School of Physics and Astronomy, Monash University, Clayton, Victoria 3800, Australia}
\author[0000-0003-2611-7269]{Keiichi Maeda}\affiliation{Department of Astronomy, Kyoto University, Kitashirakawa-Oiwake-cho, Sakyo-ku, Kyoto 606-8502, Japan}

\begin{abstract}
The red supergiant (RSG) problem refers to the observed dearth of luminous RSGs identified as progenitors of Type II supernovae (SNe II) in pre-SN imaging. Understanding this phenomenon is essential for studying pre-SN mass loss and the explodability of core-collapse SNe. In this work, we re-assess the RSG problem using late-phase spectroscopy of a sample of 50 SNe II. The [O I] $\lambda\lambda$6300,6363 emission in the spectra is employed to infer the zero-age main sequence (ZAMS) mass distribution of the progenitors, which is then transformed into a luminosity distribution via an observation-calibrated mass-luminosity relation. The resulting luminosity distribution reveals an upper cutoff at $\log\,L/L_{\odot}\,=\,5.21^{+0.09}_{-0.07}\,$dex, and the RSG problem is statistically significant at the 2$\sigma$ to 3$\sigma$ level. Assuming single RSG progenitors that follow the mass-luminosity relation of \texttt{KEPLER} models, this luminosity cutoff corresponds to an upper ZAMS mass limit of $20.63^{+2.42}_{-1.64}\,M_{\odot}$. Comparisons with independent measurements, including pre-SN imaging and plateau-phase light curve modeling, consistently yield an upper ZAMS mass limit below $\sim$\,25\,$M_{\rm \odot}$, with a significance level of 1–3$\sigma$. While each individual method provides only marginal significance, the consistency across multiple methodologies suggests that the lack of luminous RSG progenitors may reflect a genuine physical problem. Finally, we discuss several scenarios to account for this issue should it be confirmed as a true manifestation of stellar physics.
\end{abstract}

\section{Introduction} \label{sec:intro}
When massive stars, with zero-age-main-sequence (ZAMS) mass $M_{\rm ZAMS}$\,\( > \)\,8\,$M_{\rm \odot}$ exhaust the nuclear fuel in their core, they will experience iron-core infall, and explode as a core-collapse (CC) supernova (SN). The CCSNe are classified based on the absorption features that emerge in the early phase spectroscopy; those with hydrogen lines, therefore probably possessing a massive hydrogen-rich envelope, are classified as type II supernovae (SNe II), while those without hydrogen lines are classified as stripped-envelope supernovae (SESNe). The readers may refer to \citet{filippenko97,galyam17,modjaz19} for the classification of CCSNe.

The first SN II that have directly confirmed red supergiant (RSG) progenitor from pre-supernova imaging was SN 2003gd \citep{2003gd}, whose spectral energy distribution (SED) was consistent with those of field RSGs. As the sample of SNe II with directly identified progenitors from pre-SN imaging has grown, it has become increasingly evident that most SNe II originate from RSG progenitors, as predicted by stellar evolution theory. However, a discrepancy has emerged: while the most luminous field RSGs can have bolometric luminosity reaching log\,$L/L_{\rm \odot}$\,=\,5.5\,dex (\citealt{DCB18})\footnote{Throughout this work, log\,$L$ refers to bolometric luminosities and are expressed in solar units unless otherwise specified.}, no evidence exists for such bright RSGs as progenitors of SNe II. Indeed, the most luminous RSG progenitor detected to date is that of SN 2009hd, with log\,$L$\,=\,5.24\,dex, or $M_{\rm ZAMS}$\,$\sim$\,20\,$M_{\rm \odot}$, although the conversion from luminosity to $M_{\rm ZAMS}$ is model dependent. This apparent absence of bright and massive RSG progenitors for SNe II is referred to as the RSG problem (\citealt{smartt09, walmswell12, eldridge13, meynet15, smartt15, DB18, strotjohann24}).

This observational discrepancy challenges stellar evolution theory. According to single-star models, only stars with $M_{\rm ZAMS}$\,\( > \)\,30\,$M_{\rm \odot}$ are predicted to experience sufficiently strong stellar winds to completely strip their hydrogen-rich envelopes (\citealt{meynet00,sukhbold16}), and one would expect to observe SNe II with $M_{\rm ZAMS}$ between 20 and 30\,$M_{\rm \odot}$. Several theories have been proposed to explain the absence of RSG progenitors within this range. One possibility is the failed SN scenario, which suggests that RSGs within this mass range collapse to form black holes and disappear quietly without producing a bright explosion (\citealt{oconnor11,horiuchi14,pejcha15,ertl16,muller16,sukhbold16,sukhbold18,ebinger19,sukhbold20,fryer22,temaj24}). Another explanation involves eruptive mass loss, where instabilities in RSGs of this mass range lead to significant mass ejections, stripping their hydrogen-rich envelopes. Such stars may instead explode as SESNe or interacting SNe, rather than SNe II (\citealt{smith09,yoon10_pulse,SA14,meynet15,temaj24}).

Despite these theoretical investigations, efforts have been made to assess the significance of the RSG problem. Converting pre-SN magnitudes to bolometric luminosities depends on several assumptions, such as the spectral type of the progenitor, circumstellar dust properties (\citealt{walmswell12,vandyk24}), and bolometric corrections (\citealt{DB18,healy24,vandyk24,beasor25}), all of which introduce substantial uncertainties. Recently, \citet{healy24} and \citet{beasor25} demonstrated that using a single bandpass for progenitor identification, as was done for many RSGs, can lead to systematic underestimations of luminosities, and found no statistical significant evidence of missing high luminosity RSGs in pre-SN images. Statistical limitations also play a role; \citet{DB18} argued that the RSG problem might be partly due to the small sample size of observed RSG progenitors. Additionally, \citet{strotjohann24} raised concerns about the impact of telescope sensitivity on RSG progenitor detection statistics.

Given these considerations, it is important to investigate other methodologies to infer $M_{\rm ZAMS}$ independently to cross-check the significance level of the RSG problem. One of the most frequently adopted techniques is modeling the light curve at plateau phase (\citealt{morozova18,martinez22,moriya23}). This method involves evolving models with different $M_{\rm ZAMS}$ until the onset of core-collapse and then injecting different amounts of energy and $^{56}$Ni into the central region to trigger the explosion. The resultant light curve is compared with observation to determine these quantities. By employing \texttt{KEPLER} models as RSG progenitors, \citet{morozova18} found an upper $M_{\rm ZAMS}$ cutoff at 22.9\,$M_{\rm \odot}$ for a sample of 20 SNe II. In a similar investigation on a larger sample, \citet{martinez22} found the upper cutoff at 21.3\,$M_{\rm \odot}$. This approach has the advantage of allowing for relatively large samples, however, it also has limitations: the properties of the plateau light curve are mainly determined by the mass of the hydrogen-rich envelope $M_{\rm Henv}$ (when other properties, such as the explosion energy and the radius of the RSG, are fixed) rather than $M_{\rm ZAMS}$ itself \citep{kasen09,dessart19,goldberg19,hiramatsu21,fang24a}. The validity of this approach depends on the assumption of a unique relationship between $M_{\rm ZAMS}$ and the envelope mass, which may hold for single stars but can break down in the presence of a binary companion (\citealt{heger03, eldridge08, yoon10, smith11, sana12, smith14, yoon15, yoon17, ouchi17, eldridge18, fang19, zapartas19, zapartas21, chen23, ercolino23, fragos23, hirai23, matsuoka23, sun23}, among many others) or uncertainties in stellar winds (\citealt{eldridge06,mauron11,meynet15,DB18,DB20,wang21,massey23,vink23,yang23,zapartas24}, among many others).   

In this work, we investigate the RSG problem using late-phase ($nebular$ phase) spectroscopy of SNe II, taken on $\sim$\,200 days after the explosion. During this phase, the spectroscopy is dominated by emissions lines, of particular importance is the oxygen emission [O I] $\lambda\lambda$6300,6363. The [O I] emission is considered as an important tool for measuring the oxygen content in the ejecta (\citealt{fransson89,maguire12,jerk12,jerk14,kuncarayakti15,SNDB,dessart20,dessart21,fang22}), which is monotonically dependent on $M_{\rm ZAMS}$ and therefore the luminosity of the RSG progenitor \citealp{sukhbold18,takahashi23}. As a result, the $M_{\rm ZAMS}$ inferred from the strength of the [O I] line can be considered as an independent view point on the RSG problem from pre-SN images.

This paper is organized as follows: In \S 2,we describe the nebular spectroscopy sample and the methods used to process them. In \S 3, we introduce the method to determine $M_{\rm ZAMS}$ for individual SNe from [O I] emission, and establish the $M_{\rm ZAMS}$ distribution of the full sample. In \S 4, we correlate the $M_{\rm ZAMS}$, determined in \S 3, with the luminosities of the RSG progenitors from pre-SN images for a sub-sample of SNe II. This calibrated mass-luminosity relation is applied to the full sample to establish the luminosity distribution of their RSG progenitors, which is modeled with a power law function in \S5 to assess the significance of the RSG problem. We discuss the physical implications in \S6. Finally, we summarize our conclusions in \S7.

\section{Nebular spectroscopy processing}
\begin{figure*}
\epsscale{1.1}
\plotone{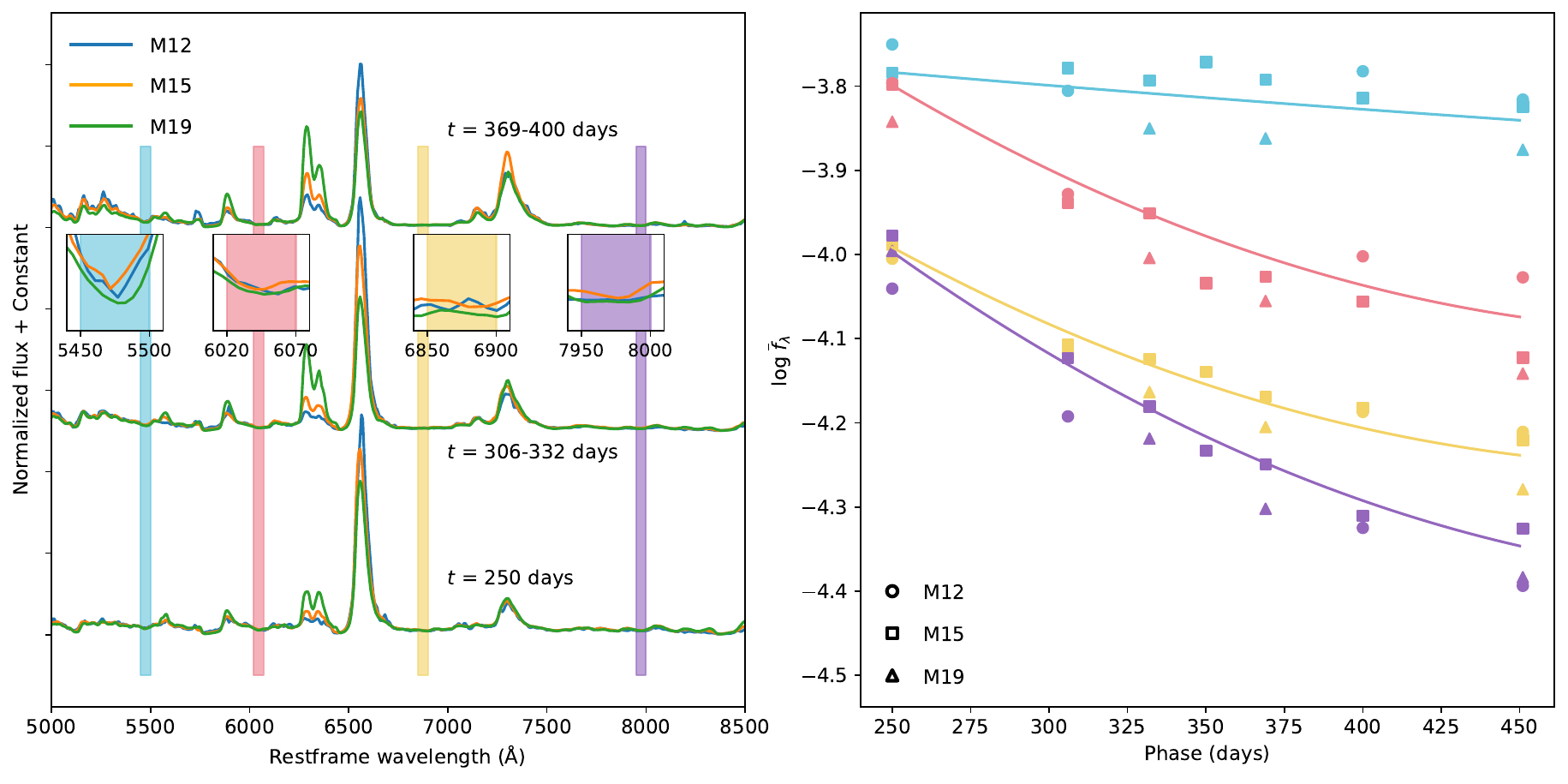}
\centering
\caption{Left panel: The nebular spectra from \citet{jerk12}, normalized to the integrated fluxes. The shaded regions mark the wavelength ranges that employed to calibrate the background fluxes. Right panel: The average fluxes in the colored regions as functions of time. The solid lines are the quadratic fits to the data points.}
\label{fig:background_model}
\end{figure*}

In this work, we compile nebular spectroscopy of SNe II from the literature that meets the following criteria: (1) that the wavelength range must cover 5000 to 8500 $\rm{\AA}$, (2) that the spectra must be obtained more than 200 days after the explosion to ensure the nebular phase is reached, but not later than 450 days to allow for comparison with spectral models; (3) that the spectra are available on the Open Supernova Catalog (\citealt{open_SN}), the Weizmann Interactive Supernova Data Repository (WISeREP; \citealt{wiserep}) or the Supernova Database of UC Berkeley (SNDB; \citealt{sndb_origin}). For objects with multiple nebular spectra available, we select the one closest to 350 days post-explosion. This phase is chosen because it is late enough to ensure all SNe II are fully nebular, yet not so late that flux contributions from shock-circumstellar material (CSM) interaction become significant (\citealt{dessart21,R22,dessart23}). The final sample consists of 50 SNe II, which are listed in Table~\ref{tab:SNe sample} in the Appendix. While this sample does not encompass all SNe II nebular spectra in the literature, a size of $N\,=\,$50 is sufficient for statistical analysis.

The absolute or relative strengths of the [O I] emission line that emerges in the nebular spectroscopy of SNe II are useful indicators of the carbon-oxygen (CO) core mass and, consequently, the ZAMS mass of the progenitor. In this work, we use the fractional flux of the [O I] line within the wavelength range of 5000 to 8500 $\rm{\AA}$, $f_{\rm [O\,I]}$, as a diagnostic for the oxygen mass in the ejecta. We compare these measurements with model spectra from \cite{jerk12} and \cite{jerk14}, following a methodology similar to that of \cite{barmentloo24} and \citet{dessart21}. The wavelength range is chosen to encompasses all the observed spectra in the sample, and cover most of the main emission features in the optical band.
This approach has an important advantage: because $f_{\rm [O\,I]}$ measures relative fluxes, it is unaffected by distance and flux calibration, and is insensitive to extinction in the host environment as long as it is not highly extincted, which typically constitutes one of the largest sources of uncertainty. However, to measure $f_{\rm [O\,I]}$ and make a meaningful comparison with the models, the observed spectra must first be standardized, as described below.

The nebular spectra of SNe II consist of multiple prominent emission lines, including [O I] $\lambda\lambda$6300,6363, H$\alpha$, and [Ca II] $\lambda\lambda$7291,7323, superimposed on a so-called pseudo-continuum formed by thousands of weak spectral lines. Figure~\ref{fig:background_model} shows the model spectra of SNe II taken from \cite{jerk12}, normalized to their integrated flux within the wavelength range 5000–8500 $\rm{\AA}$. Hereafter, we refer to models from $M_{\rm ZAMS}$\,=\,12\,$M_{\rm \odot}$ as M12 models, while those from $M_{\rm ZAMS}$\,=\,15\,$M_{\rm \odot}$ and 19\,$M_{\rm \odot}$ are referred to as M15 and M19 models, respectively. In the left panel of Figure~\ref{fig:background_model}, four spectral regions are specifically highlighted: 5450–5500, 6020–6070, 6850–6900, and 7950–8000\,${\rm \AA}$. These wavelength ranges do not contain strong emission lines, allowing the fluxes in these regions to be treated as pure pseudo-continuum \citep{barmentloo24}.
As shown in the right panel of Figure~\ref{fig:background_model}, the average fluxes $\overline{f}({\lambda}_{\rm i},t)$ within these regions are independent of $M_{\rm ZAMS}$ but well determined by the spectral phase $t$, which are fitted by quadratic functions (the solid lines in the right panel of Figure~\ref{fig:background_model}) to estimate $\overline{f}({\lambda}_{\rm i},t)$ at arbitrary phases. This forms the basis for the approach to addressing contamination from the host environment. 

\begin{figure*}
\epsscale{1.15}
\plotone{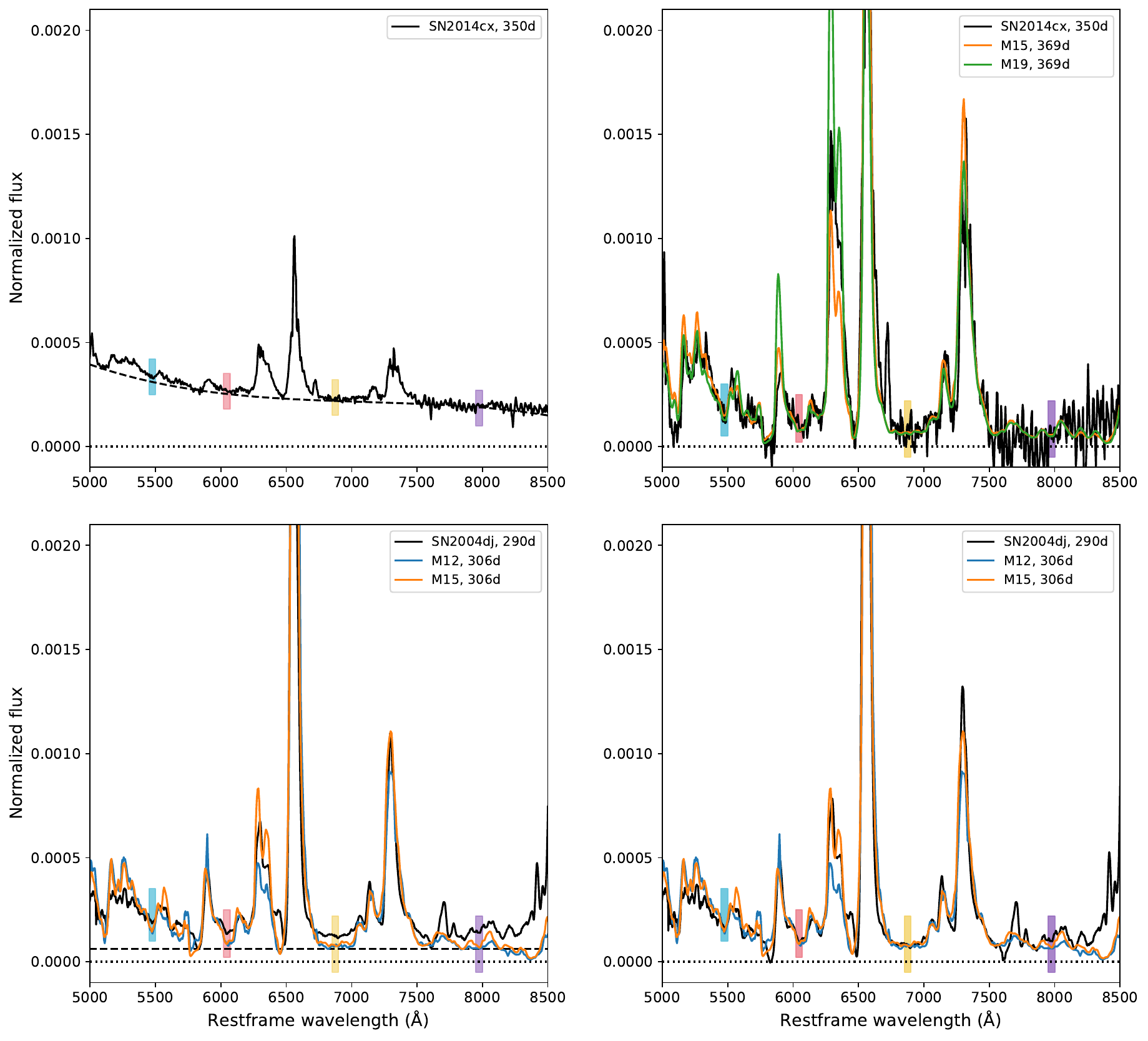}
\centering
\caption{Examples of the removal the possible contamination from the host galaxy. Upper panels: SN 2014cx; lower panels: SN 2004dj. In the left panels, the uncalibrated spectra are normalized to the integrated flux within 5000 to 8500\,$\rm \AA$. The colored strips highlight the wavelength regions used to determine and subtract the contamination flux $f_{\rm con}$ from the host galaxy, which is represented by the black dashed lines. A quadratic form (upper) and constant (lower) form of $f_{\rm con}$ are applied. Right panels: The spectra normalized to integrated flux within 5000 to 8500\,$\rm \AA$, after $f_{\rm con}$ are removed. Models from \citet{jerk12} are overlaid for comparison, showing good agreement between the fluxes in the colored regions and the models. The dotted line in all panels represents zero flux.}
\label{fig:background_subtraction}
\end{figure*}

The observed spectra of SNe 2014cx and 2004dj, normalized to their integrated flux within the wavelength range 5000–8500 $\rm{\AA}$, are illustrated in the left panels of Figure~\ref{fig:background_subtraction}. The spectrum of SN 2014cx shows significant contamination from its host environment, as indicated by its unusual slopes, similar to SN 2012ec \citep{jerk15}. Although the case of SN 2004dj is less extreme, the average fluxes in the aforementioned wavelength ranges consistently exceed those predicted by the spectral models at the same phase. This discrepancy suggests that the background emission might not have been completely removed during the processing of the raw observational data. Before measuring $f_{\rm [O\,I]}$, these residual fluxes are removed as follows:
\begin{enumerate}
    \item the observed flux $f_{\rm obs}$ in the rest frame are transformed to the standardized (or normalized) flux $f_{\rm norm}$ assuming:
\begin{equation} \label{eq:transform}
f_{\rm norm} = A\,\times\,(f_{\rm obs} - f_{\rm con}),
\end{equation}
here $f_{\rm con}$ is the fluxes of the residual from the host, and $A$ is a normalized constant. The destination function, normalized flux $f_{\rm norm}$, should meet the requirement that when normalized to unity, the fluxes from the aforementioned 4 regions should be close to the model spectra at the same phase, which are estimated from the quadratic fits in the right panel of Figure~\ref{fig:background_model}. For most SNe in the sample, $f_{\rm con}$ is assumed to be a constant for simplicity. However, for 3 objects (SNe 2012ch, 2012ec, and 2014cx) that exhibit unusual spectral slopes due to the contamination of their bright host environments, a quadratic form of $f_{\rm con}$ is applied:
\begin{align*}
f_{\rm con} = b_{\rm 2}\,\lambda^2 + b_{\rm 1}\,\lambda + b_{\rm 0},
\end{align*}
where $\lambda$ is the wavelength. 

    \item The \texttt{Python} package \texttt{scipy.optimize} is imported to find the pair \{$A$,\,$f_{\rm con}$\} (or \{$A$,\,$b_{\rm 0}$,\,$b_{\rm 1}$,\,$b_{\rm 2}$\} for $f_{\rm con}$ in quadratic form) that minimize the following quantities:
\begin{align*}
r_{\rm 0} &= \int_{\rm 5000\,\AA}^{{\rm 8500\,\AA}} f_{\rm norm}~d\lambda - 1\\
r_{\rm i} &=  \overline{f}_{\rm norm}({\lambda}_{\rm i}) - \overline{f}_{\rm model}({\lambda}_{\rm i}, t),~i = 1,2,3,4.\\
\end{align*}
Here $\overline{f}_{\rm norm}({\lambda}_{\rm i})$ is the average fluxes of the observed spectra after normalization within the 4 selected wavelength ranges, and $\overline{f}_{\rm model}({\lambda}_{\rm i},t)$ is the average pseudo-fluxes of the spectral models at the same phase $t$, estimated from the quadratic fit (the solid lines in the right panel of Figure~\ref{fig:background_model}). These procedures ensure that, after the transformation described by Equation~\ref{eq:transform}, the integral of the normalized flux within 5000 to 8500\,${\rm \AA}$ equals unity. Moreover, the fluxes of the pseudo-continuum within the specific regions are aligned with those of the models, allowing for fair comparison.
\end{enumerate}

The above procedures are applied to SN 2014cx and 2004dj, where quadratic and constant forms of $f_{\rm con}$ are assumed respectively, and the resultant $f_{\rm con}$ are shown as the black dashed lines in the left panels of Figure~\ref{fig:background_subtraction}. The normalized fluxes, after $f_{\rm con}$ is removed, are shown in the right panels and compared with the spectral models at similar phases, showing that the pseudo-continuum fluxes of the normalized observed spectra are consistent with those of the models. After the spectra is normalized to $f_{\rm norm}$, we fit the lines in the range of 6100 to 6800\,${\rm \AA}$ with multi-Gaussian functions: (1) two Gaussians with the same standard deviation and peaks separated by 63\,${\rm \AA}$ to represent the [O I] doublet, (2) one Gaussian centered near 6563\,${\rm \AA}$ (with a small allowed shift) to represent H$\alpha$, and (3) an additional Gaussian with an arbitrary center to account for a spectral feature commonly observed between [O I] and H$\alpha$ in many SNe II nebular spectra. The fluxes of [O I] and H$\alpha$ are measured by integrating the fitted profiles. Figure~\ref{fig:background_model} illustrates this fitting procedure using SNe 2014G and 2023ixf as examples. Although SN 2023ixf exhibits a complex [O I] line profile (see, e.g. \citealt{2023ixf_nebular,fang24b}), likely reflecting intricate ejecta geometry (\citealt{fang24}), this study focuses solely on the integrated flux, and these complexities are not considered.

\begin{figure}[!htb]
\epsscale{1.1}
\plotone{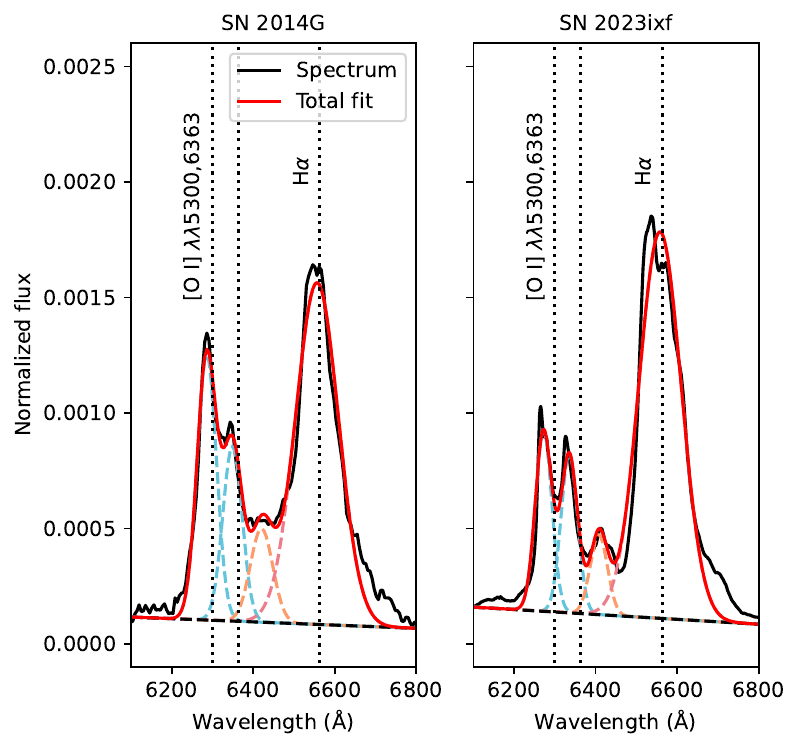}
\centering
\caption{The fitting to the line profiles of SNe 2014G (left) and 2023ixf (right). The black solid line is the normalized observed spectrum and the black dashed line is the psuedo-continuum. The light blue, pink and orange dashed lines represent the fits to the [O I] $\lambda\lambda$6300,6363, H$\alpha$ and the sub-structure respectively. The red solid line is the sum of the fitted profiles.}
\label{fig:fit_example}
\end{figure}

The uncertainties in the fractional flux of [O I] (H$\alpha$), $f_{\rm [O\,I]}$ ($f_{{\rm H}\alpha}$), come mainly from uncertainties in subtracting the contamination flux $f_{\rm con}$. This is quantified using a Monte Carlo method: the original observed spectra are first smoothed, and the fluxes in the four fitting regions are replaced with the smoothed fluxes, augmented by random noises. The noise level in each region is estimated as the standard deviation of the difference between the original flux and the smoothed flux within it. The determination of $f_{\rm con}$ and the measurement of the [O I] (or H$\alpha$) fluxes are then repeated 1000 times, following the same procedure. In each trial, the pseudo-continuum fluxes are allowed to randomly vary by at most 20\% (0.08 dex; see the scatter lever in the right panel of Figure~\ref{fig:fit_example}). The standard deviations of these measurements are taken as the uncertainties in the fractional line fluxes.

In addition to $f_{\rm [O\,I]}$, we further define its regulated form as
\begin{align*}
f_{\rm [O\,I],reg} =\,\frac{f_{\rm [O\,I]}}{1 - f_{\rm H\alpha}},
\end{align*}
i.e., the fractional flux of [O I] after the H$\alpha$ line is subtracted from the spectrum. The motivation for introducing $f_{\rm [O\,I],reg}$ is to address the uncertainties associated with H$\alpha$ arising from various factors, which will be discussed in \S3.

\section{Estimation of ZAMS mass}
\begin{figure}[!htb]
\epsscale{1.1}
\plotone{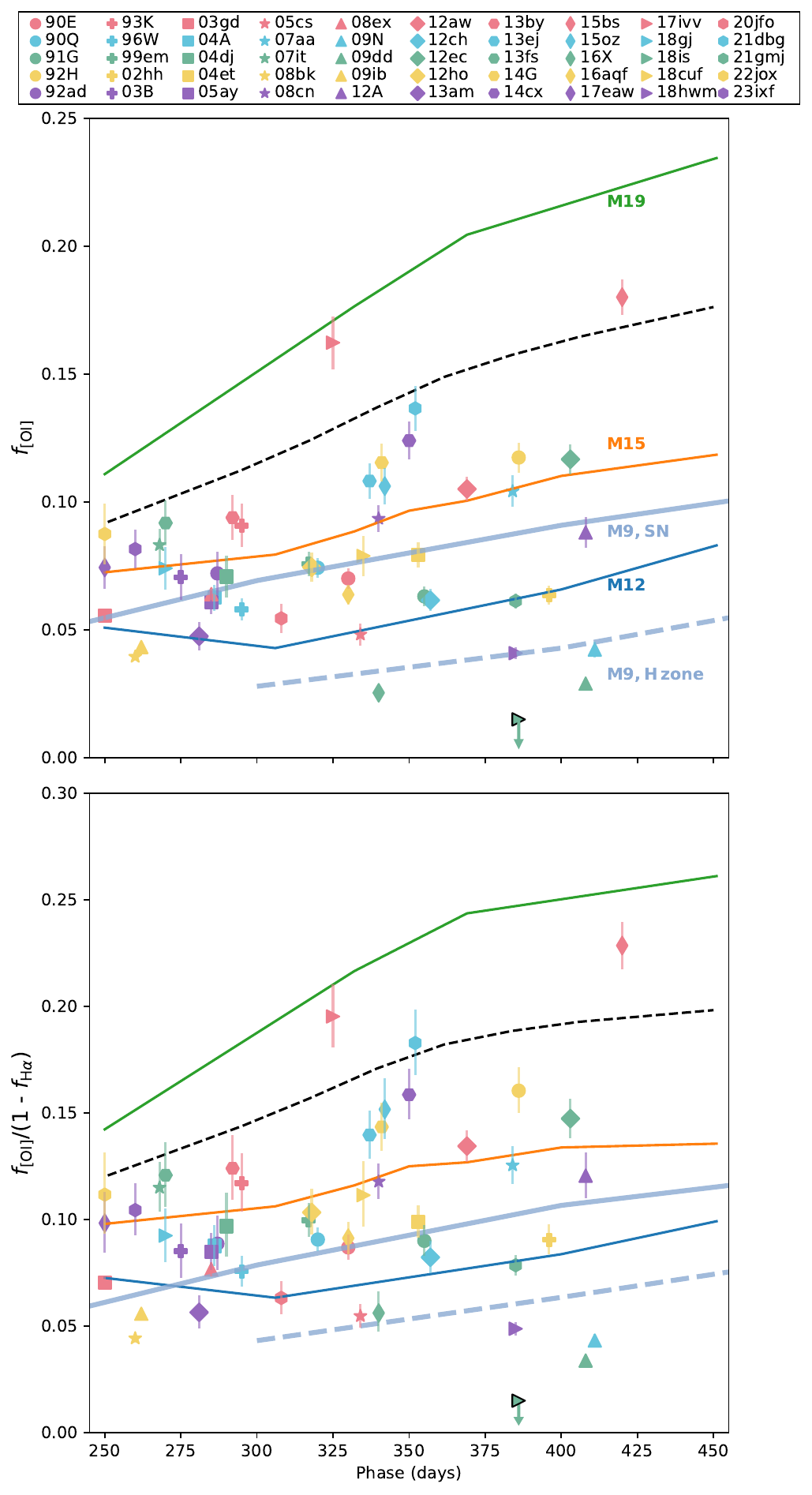}
\centering
\caption{Upper panel: The fractional fluxes of [O I] of individual SNe (labeled by different colors and markers) compared with the model tracks. The black dashed line is the average of the M15 and M19 tracks that represent the case when $M_{\rm ZAMS}$\,=\,17\,$M_{\rm \odot}$. Lower panel: same as upper panel, but for cases of regulated fractional fluxes of [O I] (see main text).}
\label{fig:track}
\end{figure}

In the upper panel of Figure~\ref{fig:track}, the measured fractional fluxes of [O I], $f_{\rm [O\,I]}$, are compared with the spectral models from \citet{jerk12}. A significant proportion of the SNe in the sample cluster between the M12 and M15 tracks. Compared to the results of \citet{valenti16}, the size of the nebular spectroscopy sample is substantially larger, and more objects are found to lie above the M15 tracks. However, no single object provides evidence for a progenitor more massive than the M19 model. Notably, with the exception of SNe 2015bs (\citealt{2015bs}) and 2017ivv (\citealt{2017ivv}), no individual SN in the sample has a derived $M_{\rm ZAMS}$ exceeding 17\,$M_{\rm \odot}$. 

The fractional fluxes of the M9 models from \citet{jerk18}, computed from the progenitor model with $M_{\rm ZAMS}$\,=\,9\,$M_{\rm \odot}$ using \texttt{KEPLER} in \citet{sukhbold16}, are plotted for reference. These models include two sets: one representing the total SN spectrum (M9, SN) and another where the core material is replaced with H-zone material (M9, H-zone). For the former case, the fractional [O I] flux exceeds the M12 track, whereas in the latter, it falls below the M12 track. Despite having a lower oxygen mass, $f_{\rm [O\,I]}$ of the M9 models are not consistently lower than the M12 models.

This non-monotonic behavior may stem from several factors. First, the M9 progenitor undergoes a strong silicon flash, which could alter its core properties compared to the M12, M15, and M19 models, where nuclear burning proceeds stably throughout evolution (\citealt{sukhbold16}). Furthermore, these models assume lower explosion energies, leading to relatively narrow emission lines with a full width at half maximum (FWHM) of $\sim$\,1000 km\,s$^{-1}$, a feature not commonly observed in most SNe II. 

Given these unique characteristics, the non-monotonic [O I] flux behavior, and uncertainties regarding whether the full SN models or pure H-zone models should be applied, the M9 models are not used to refine $M_{\rm ZAMS}$ estimates below the M12 track. Notably, SNe 2005cs, 2008bk, and 2018is, which were compared with M9 models in \citet{jerk18}\footnote{For SN 1997D, the wavelength range and the spectral phase do not meet the criteria in this work} and \citet{2018is}, exhibit fractional [O I] fluxes below the M12 track, suggesting that the M12 models already capture the low-mass nature of their progenitors.

The comparison of $f_{\rm [O\,I],reg}$ and the spectral models are shown in the lower panel of Figure~\ref{fig:track}. The motivation for the introduction of $f_{\rm [O\,I],reg}$, i.e., removing H$\alpha$ from $M_{\rm ZAMS}$ estimates, stems from the fact that this line can be affected by many factors as will be discussed below.

Growing evidences suggest that some SNe II experienced partially stripping of the hydrogen-rich envelope prior to their explosions. A reduced envelope mass $M_{\rm Henv}$ is necessary to explain the short duration light curves observed in several individual SNe, such as SNe 2006Y, 2006ai, 2016egz (\citealt{anderson14,hiramatsu21}), 2018gj (\citealt{2018gj}), 2020jfo (\citealt{2020jfo2}), 2021wvw (\citealt{2021wvw}), 2023ixf (\citealt{fang24b,hsu24}) and 2023ufx (\citealt{2023ufx1, 2023ufx2}). \citet{fang24a} further suggest that, to account for the observed diversity in SNe II light curves (e.g., \citealt{anderson14,valenti16,anderson24}), approximately half of SNe II must have stripped their envelopes to about $\sim$\,4.0\,$M_{\rm \odot}$ (see also \citealt{hiramatsu21}). 
    
The spectral models from \cite{jerk12} assume a massive hydrogen-rich envelope, and therefore may not accurately reproduce the H$\alpha$ flux if the hydrogen-rich envelope is partially removed. Although the exact relationship between H$\alpha$ flux and $M_{\rm Henv}$ has not yet been fully established, a pioneering study by \citet{dessart20} shows that for SNe II with $M_{\rm Henv}$ \( < \) 3\,$M_{\rm \odot}$, the H$\alpha$ flux decreases dramatically, while other parts of the spectrum are much less affected (see also \citealt{2023ufx2}). From observation, SNe II with low $M_{\rm Henv}$, inferred from plateau light curve modeling, also show relatively weak H$\alpha$ in nebular spectroscopy (\citealt{2020jfo2,2018gj,2023ufx2,fang24b}), and faster declines in their radiative tails during the late phases, compared to the cases with full $\gamma$-ray trapping (\citealt{anderson14,gutierrez17}).

Furthermore, H$\alpha$ can be illuminated by shock-CSM interaction, a process not included in the models of \citet{jerk12}, introducing another source of uncertainty in the H$\alpha$ flux. As demonstrated in \citet{dessart23}, the contribution from shock-CSM interaction can increase the integrated flux by enhancing the H$\alpha$ line in nebular phase, and it can explain the H$\alpha$ features for several objects, including SNe 2014G and 2013by. However, since currently we are still lacking consistent interacting models of SNe II in the nebular phase, it remains challenging to quantify the size of this effect.

Given these considerations, we introduce $f_{\rm [O,I],reg}$ as a simple yet effective approach to reduce the uncertainties in H$\alpha$ by removing it from the measurement. Indeed, $f_{\rm [O,I],reg}$ and $f_{\rm [O,I]}$ represent two limiting scenarios: (1) the first scenario, based on $f_{\rm [O\,I],reg}$, assumes that the change in the integrated flux, whether due to the increase in $\gamma$-ray leakage from reduced $M_{\rm Henv}$ or the contribution from interaction power, only affects the H$\alpha$ flux, leaving other spectral features unaffected; (2) the second scenario, based on $f_{\rm [O\,I]}$, assumes that the [O I] flux scales directly with the integrated flux. In this case, if the integrated flux decreases (due to additional $\gamma$-ray leakage when $M_{\rm Henv}$ is low) or increases (due to interaction power) by 50\%, the [O I] flux is also varied by the same fraction. Taking the low $M_{\rm Henv}$ models in \citet{dessart21} and the interacting SNe II models in \citet{dessart23} as references, the actual situation likely falls between these two extremes.

\begin{figure}
\epsscale{1}
\plotone{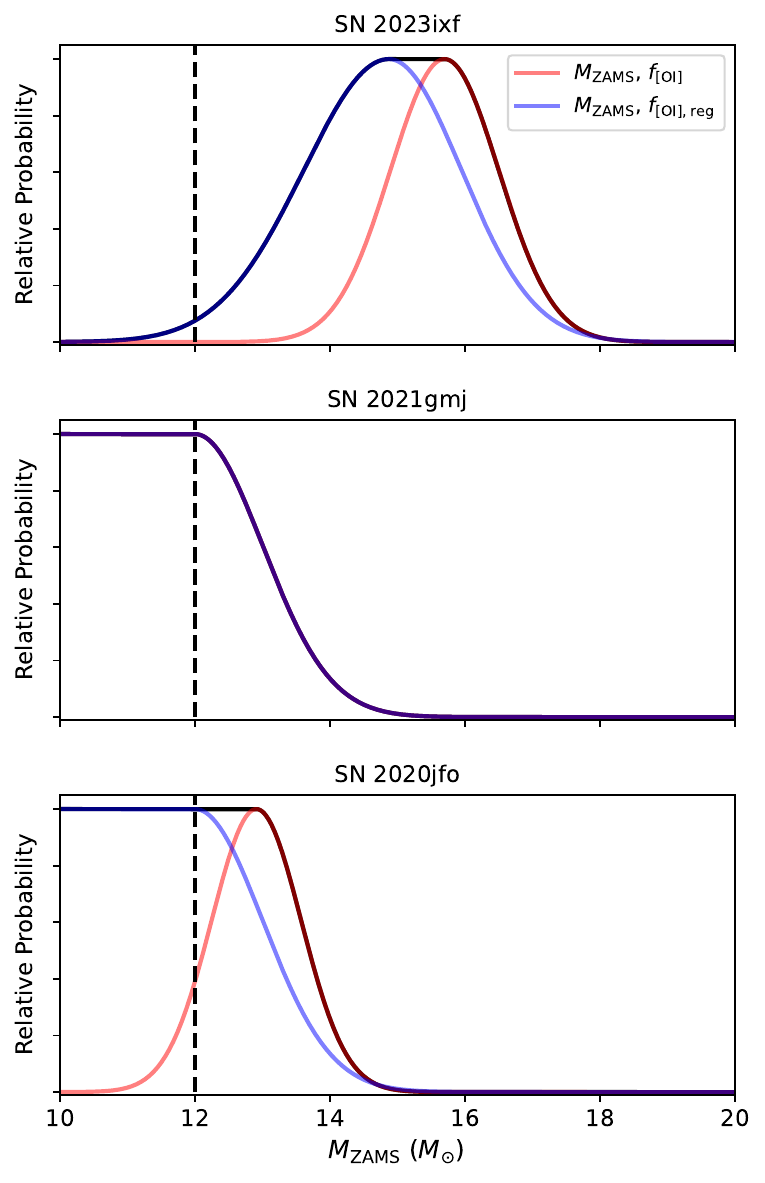}
\centering
\caption{The red and blue lines represent the measured $M_{\rm ZAMS}$ with and without the contributions from the H$\alpha$ fluxes. The final adopted $M_{\rm ZAMS}$ distributions (black lines) combine these two measurements. If the median values are above the M12 tracks, Gaussian distributions are assumed for $M_{\rm ZAMS}$ with and without H$\alpha$ fluxes, and all values between the two median values are assumed to have the same probability (SN 2023ixf; upper panel); if both measurements are below the M12 track, then $M_{\rm ZAMS}$ is assumed to be uniformly distributed between 10 and 12\,$M_{\rm \odot}$ with a Gaussian tail ($\sigma$\,=\,1\,$M_{\rm\odot}$; SN 2021gmj; middle panel); The lower panel shows the case when one of the measurement is above and the other is below the M12 track (SN 2020jfo).}
\label{fig:mass_distribution_examples}
\end{figure}

With $f_{\rm [O\,I]}$ and $f_{\rm [O\,I],reg}$, this work measures $M_{\rm ZAMS}$ for individual SNe II in the sample following the below procedures:

\begin{itemize}
    \item For $f_{\rm [O\,I]}$ above the M12 track: $M_{\rm ZAMS}$ is determined through linear interpolation with the model tracks. For example, if an SNe is observed at phase $t$ with $f_{\rm [O\,I]}$ between the M12 and M15 tracks, we first estimate $f_{\rm [O\,I]}$ of the models at $t$ through interpolation, and then interpolate between the models again to estimate $M_{\rm ZAMS}$ based on $f_{\rm [O\,I]}$;

    \item For $f_{\rm [O\,I]}$ below the M12 track: $M_{\rm ZAMS}$ is assumed to follow a uniform distribution between 10 and 12\,$M_{\rm \odot}$, with a Gaussian tail (standard deviation of 1\,$M_{\rm \odot}$) extending to higher masses.

    \item The same methodology is applied to measurements using $f_{\rm [O\,I],reg}$;

    \item The final adopted $M_{\rm ZAMS}$ combines the measurements with $f_{\rm [O\,I]}$ and $f_{\rm [O\,I],reg}$: all $M_{\rm ZAMS}$ values within the range defined by the $M_{\rm ZAMS}$ measured from $f_{\rm [O\,I]}$ and $f_{\rm [O\,I],reg}$ are assumed to be uniformly distributed. Figure~\ref{fig:mass_distribution_examples} shows 3 examples of the final adopted $M_{\rm ZAMS}$ distributions for individual SNe. For individual SNe II, its $M_{\rm ZAMS}$ is not a single value with Gaussian uncertainty but follows an irregular distribution that cannot be described analytically. Throughout this work, $M_{\rm ZAMS}$ and its uncertainty represent the median value and the 68\% confidence interval (CI; determined by the 16th and 84th percentiles) of this distribution.

\end{itemize}

\begin{figure}[!htb]
\epsscale{1.}
\plotone{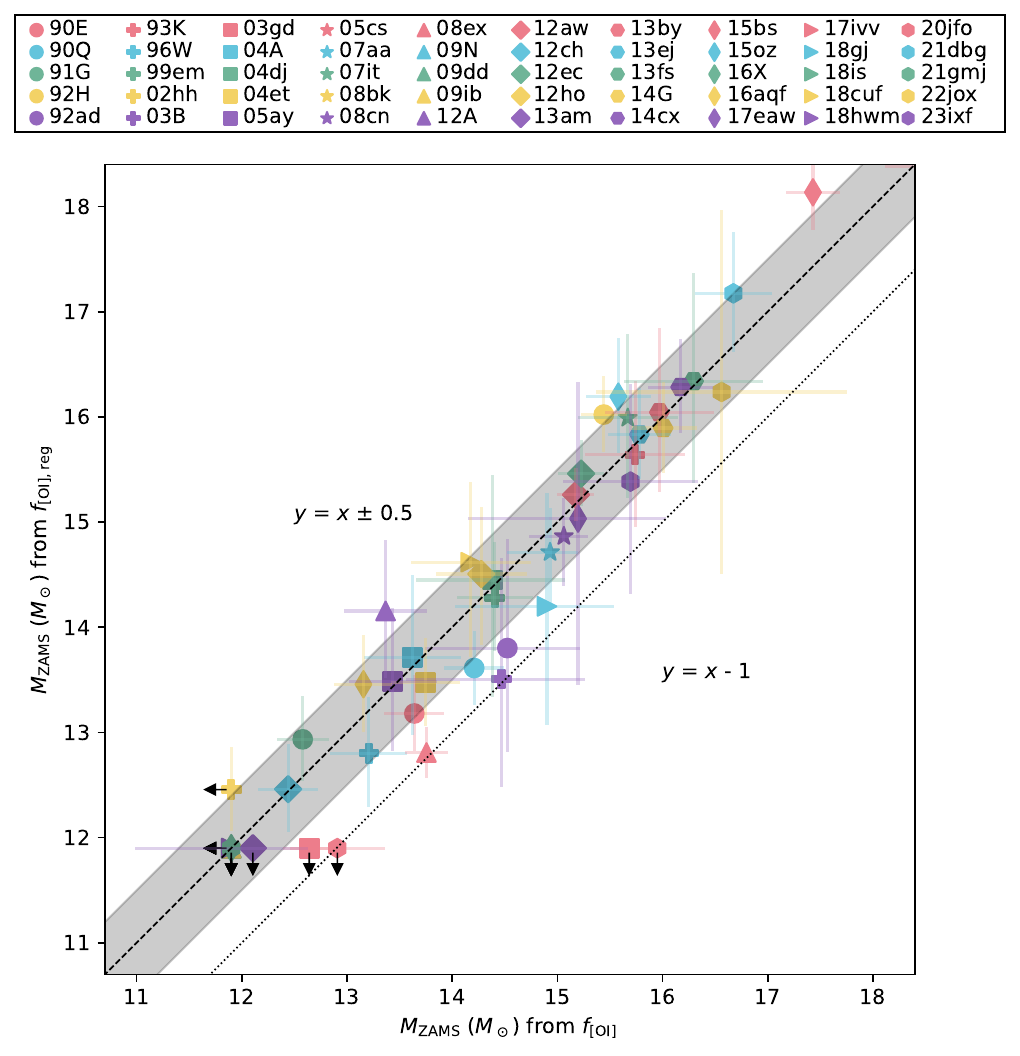}
\centering
\caption{Comparison between the $M_{\rm ZAMS}$ measured with ($f_{\rm [O\,I]}$) and without ($f_{\rm [O\,I],reg}$) the contributions from the H$\alpha$ flux. The dashed line represents $y\,=\,x$. The shaded region represent the case when the difference between the two measurements are within 0.5\,$M_{\rm \odot}$. The dotted line represent the case when $M_{\rm ZAMS}$ measured from $f_{\rm [O\,I],reg}$ is 1.0\,$M_{\rm \odot}$ smaller than that measured from $f_{\rm [O\,I]}$.}
\label{fig:compare_mass_wo_h}
\end{figure}

In Figure~\ref{fig:compare_mass_wo_h}, we compare the $M_{\rm ZAMS}$ estimated from $f_{\rm [O\,I]}$ and $f_{\rm [O\,I],reg}$. The result shows that $f_{\rm [O\,I],reg}$ generally predicts lower $M_{\rm ZAMS}$, with differences reaching up to $\sim\,1\,M_{\rm \odot}$ in some cases. This suggests that spectral models may overestimate the H$\alpha$ flux.

\begin{figure}
\epsscale{1.1}
\plotone{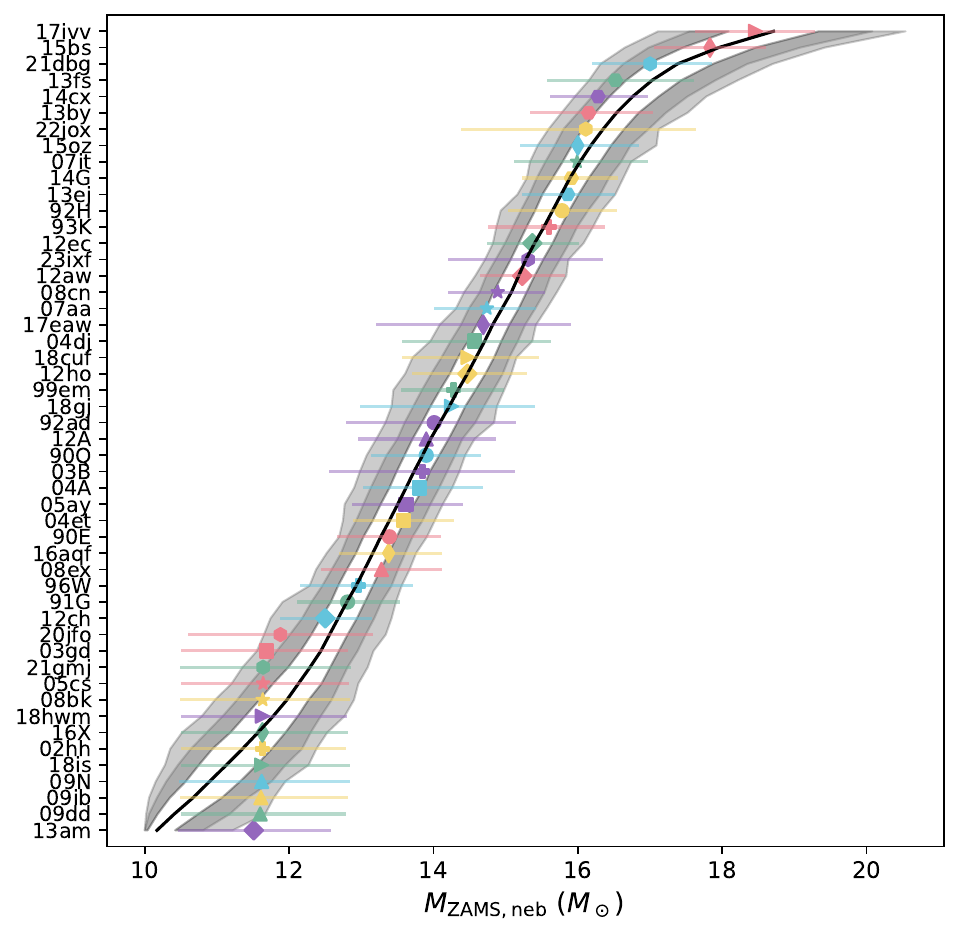}
\centering
\caption{The cumulative distribution of $M_{\rm ZAMS}$ measured from nebular spectra (denoted as $M_{\rm ZAMS,neb}$). The black solid line represent the median values for each rank from the sorted method described in the main text. For illustration purposes, the 68\% CI is not colored, while the 95 and 99.75\% CI are represented by the transparent regions.}
\label{fig:neb_mass_histogram}
\end{figure}

The distribution of $M_{\rm ZAMS}$ of the full sample is shown in Figure~\ref{fig:neb_mass_histogram}, calculated using a Monte Carlo method similar to that described in \citet{DB18}: in each trial, for each individual SN, a mass is randomly sampled from its $M_{\rm ZAMS}$ distribution. For those with upper limit, a mass is randomly sampled from a distribution that is uniform between 10 to 12\,$M_{\rm \odot}$ (as shown in Figure~\ref{fig:mass_distribution_examples}). The simulated sample is then sorted. This process is repeated 10,000 times. The median $M_{\rm ZAMS}$ for each rank, from the SN with the lowest $M_{\rm ZAMS}$ to the one with the highest $M_{\rm ZAMS}$, is calculated and represented by the black line in Figure~\ref{fig:neb_mass_histogram}. The shaded regions indicate 95 and 99.7\% CI, while the 68\% CI is not filled for illustration purposes.

\section{Estimation of RSG luminosity}
In the previous section, we developed a method to estimate $M_{\rm ZAMS}$ from nebular spectroscopy (hereafter denoted as $M_{\rm ZAMS,neb}$ for clarity). The aim of this study is to assess the significance of the RSG problem, i.e., the lack of luminous RSG progenitor for SNe II. For this purpose, our next step is to convert $M_{\rm ZAMS,neb}$ into a luminosity scale, which can, in principle, be done using the $M_{\rm ZAMS}$–luminosity relation (hereafter referred to as MLR) derived from stellar evolution models. However, $M_{\rm ZAMS, neb}$ mainly reflects the oxygen content in the ejecta. Inferring $M_{\rm ZAMS}$ from nebular spectroscopy relies on the underlying relation between the synthesized oxygen mass $M_{\rm O}$ and $M_{\rm ZAMS}$. While $M_{\rm O}$ is a monotonic function of $M_{\rm ZAMS}$, with more massive stars generally synthesizing more oxygen, the transformation from $M_{\rm O}$ (nebular spectroscopy) to $M_{\rm ZAMS}$, and subsequently to log\,$L$ strongly depends on the microphysics of the stellar evolution code, particularly the assumptions about internal mixing. Converting $M_{\rm ZAMS,neb}$ to log\,$L$ essentially reflects a $M_{\rm O}$-log\,$L$ relation, which, as will be demonstrated in the discussion in \S 4.1, is subject to significant uncertainties. To address this, an empirical MLR based on observations is established in \S4.2, and its robustness is tested in \S4.3.

\subsection{Mass-luminosity relation of stellar evolution models}
\begin{figure}
\epsscale{1.05}
\plotone{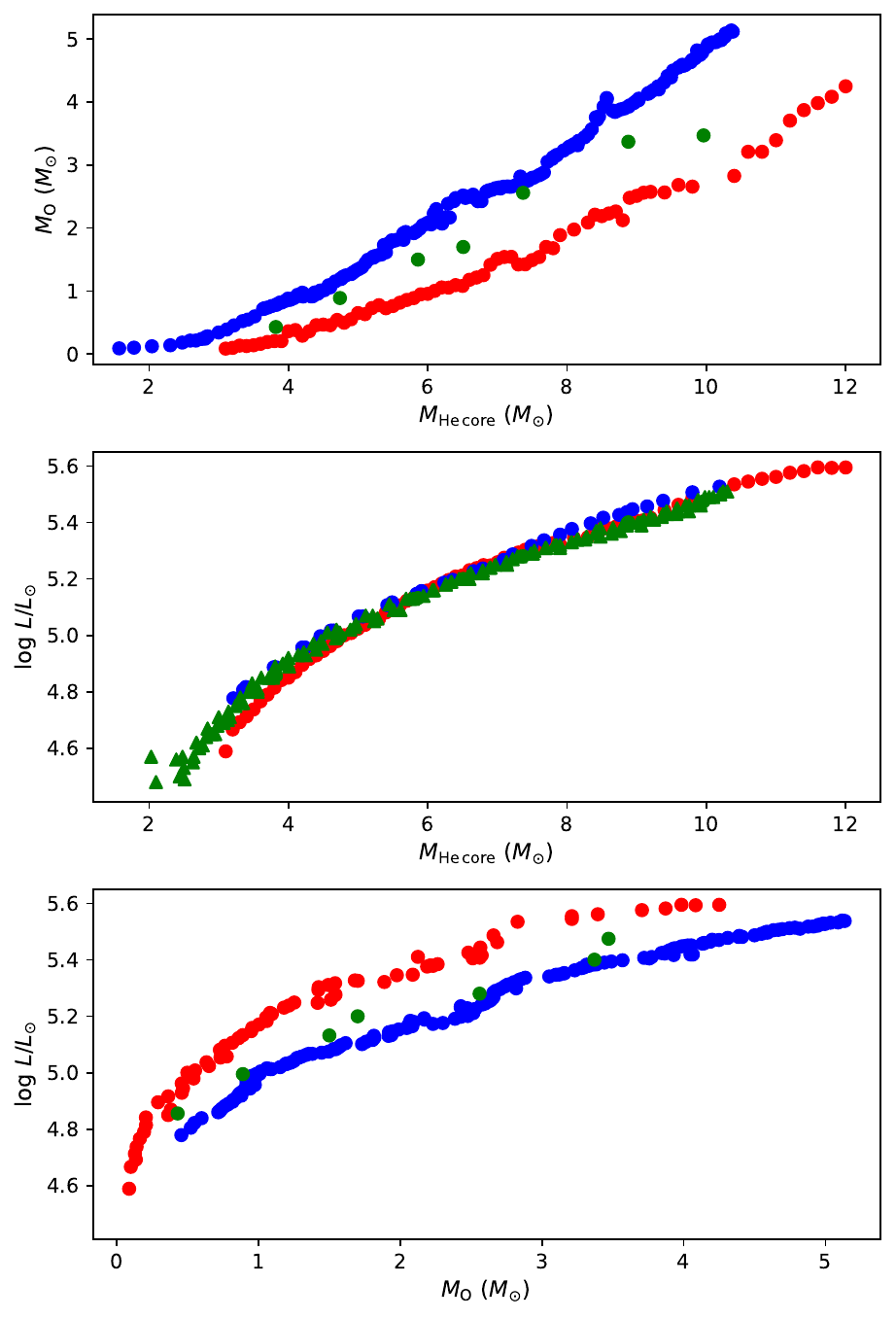}
\centering
\caption{The $M_{\rm He\,core}$-$M_{\rm O}$-log$\,L$ relations from different stellar evolution codes: \texttt{MESA} (green; \citealt{fang23,temaj24}), \texttt{KEPLER} (blue; \citealt{sukhbold16,sukhbold18,ertl20}) and \texttt{HOSHI} (red; \citealt{takahashi23}) models. Upper panel: The $M_{\rm He\,core}$-$M_{\rm O}$ relation; Middle panel: the $M_{\rm He\,core}$-log$\,L$ relation;  Lower panel: the $M_{\rm O}$-log$\,L$ relation.}
\label{fig:MHe_MCO}
\end{figure}

In the upper panel of Figure~\ref{fig:MHe_MCO}, we compare the helium core mass $M_{\rm He\,core}$ and $M_{\rm O}$ across different models to further explore these dependencies: \texttt{MESA} (progenitor models taken from \citealt{fang23}), \texttt{KEPLER} (\citealt{sukhbold16}) and \texttt{HOSHI} (\citealt{takahashi18,takahashi21,takahashi23}). We employ $M_{\rm He\,core}$ instead of $M_{\rm ZAMS}$ because $M_{\rm He\,core}$ is more directly related to the advanced nucleosynthesis once the helium core is formed after helium burning phase.

Although a clear correlation exists between $M_{\rm He\,core}$ and $M_{\rm O}$ within individual model sets, the relationship varies between different codes. Specifically, progenitors modeled with \texttt{HOSHI} produce less oxygen for a given $M_{\rm He\,core}$ compared to those from \texttt{KEPLER}, with \texttt{MESA} models lying in between. This discrepancy between the codes introduces a systematic difference of about 1.0 to 2.0\,$M_{\rm \odot}$ in the estimated $M_{\rm He\,core}$ for a given $M_{\rm O}$ (as estimated from $M_{\rm ZAMS,neb}$), which translates into approximate 2.0 to 5.0\,$M_{\rm \odot}$ difference in $M_{\rm ZAMS}$. 

In contrast, the $M_{\rm He\,core}$-log\,$L$ relation is remarkably consistent across different stellar models, as shown in the middle panel of Figure~\ref{fig:MHe_MCO}. Here we include additional \texttt{MESA} models from \citet{temaj24}. Because the models from \citet{sukhbold16} do not contain luminosity information, in this comparison, \texttt{KEPLER} models are taken from \citet{sukhbold18} and \citet{ertl20}. A linear regression returns:

\begin{equation}
{\rm log}\,\frac{L}{L_{\rm \odot}}\,=\,1.47\,\times\,{\rm log}\,\frac{M_{\rm He\,core}}{M_{\rm \odot}}\,+\,4.01,
\label{eq:Mhe_L}
\end{equation}
with the standard deviation of the residual to be 0.025 dex, equivalent to 6\% in linear scale. The tight correlation and the consistency across stellar codes indicate that, once $M_{\rm He\,core}$ is fixed, log\,$L$ can be reliably calculated using Equation~\ref{eq:Mhe_L}. \citet{schneider24} further shows that the core mass-luminosity relation is not sensitive to binarity. Hereafter, the \texttt{KEPLER} MLR refers to the transformation from $M_{\rm ZAMS}$ to $M_{\rm He\,core}$ using the relation in \citet{sukhbold16}, followed by the conversion from $M_{\rm He\,core}$ to log\,$L$ using Equation~\ref{eq:Mhe_L}. 

The lower panel of Figure~\ref{fig:MHe_MCO} illustrates the $M_{\rm O}$-log\,$L$ relation. Since the \texttt{KEPLER} models from \citet{sukhbold18} and \citet{ertl20} do not provide $M_{\rm O}$ information, we derive the $M_{\rm O}$-log\,$L$ relation for these models by combining their $M_{\rm O}$-$M_{\rm He\,core}$-log\,$L$ relation shown in the upper and middle panels of Figure~\ref{fig:MHe_MCO}. This comparison shows that, for a given $M_{\rm O}$ inferred from nebular spectroscopy, the difference in the estimated log\,$L$ can be as large as 0.2 dex. Therefore, converting $M_{\rm ZAMS, neb}$ to luminosity using MLRs from stellar evolution models can introduce significant uncertainties, and an observation-calibrated MLR is needed.

\subsection{Observation calibrated mass-luminosity relation}
\begin{figure}
\epsscale{1.1}
\plotone{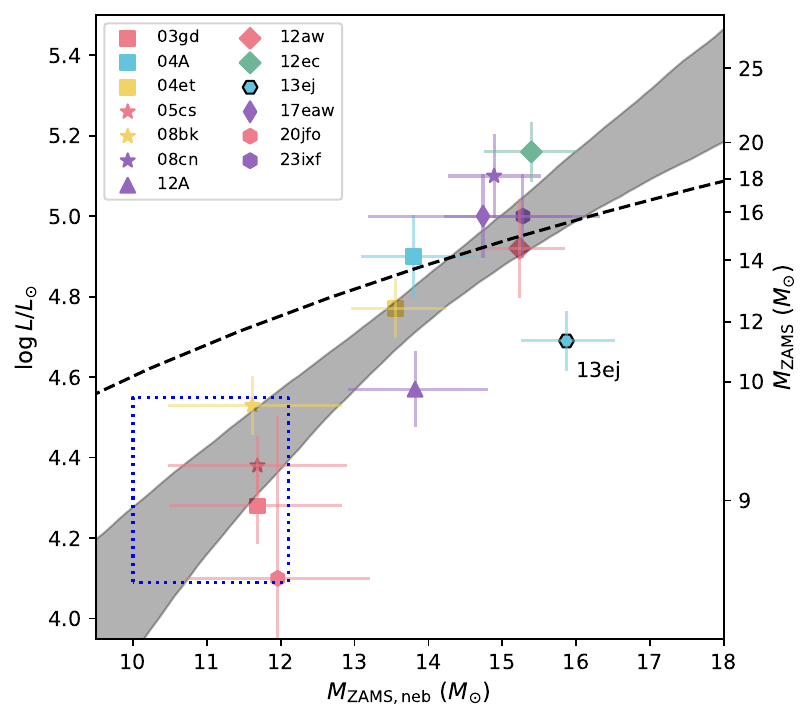}
\centering
\caption{Comparison between the $M_{\rm ZAMS,neb}$ with the luminosities of the progenitor RSGs. The shaded region is the 68\% CI of the Monte-Carlo based linear regression described in the main text, when SN 2013ej is excluded. The dashed line is the prediction if the observations follow the MLR relation of the \texttt{KEPLER} models, labeled by the right y-axis. The blue box marks the objects with $f_{\rm [O,I]}$ or $f_{\rm [O,I],reg}$ below the M12 track in Figure~\ref{fig:track}}.
\label{fig:logl_mco}
\end{figure}

To establish the $M_{\rm ZAMS, neb}$–log\,$L$ relation empirically, we use a sample of 13 SNe for which both nebular spectroscopy and RSG progenitor images are available (Figure~\ref{fig:logl_mco}). The RSG luminosities from pre-SN images log\,$L_{\rm pre\,SN}$ are mostly from \citet{DB18}, with 3 exceptions: SNe 2017eaw (\citealt{2017eaw}), 2020jfo (\citealt{kilpatrick23}) and 2023ixf (\citealt{vandyk24}), which are listed in Table~\ref{tab:preSN}. The comparison of these two quantities are shown in Figure~\ref{fig:logl_mco}.

\begin{deluxetable}{cccc}[t]
\centering
\label{tab:preSN}
\tablehead{
\colhead{SN}&\colhead{log\,$L_{\rm pre\,SN}$}&\colhead{$M_{\rm ZAMS,neb}$}&\colhead{Reference}
}
\startdata
03gd&4.28 (0.09)&11.68$^{+1.12}_{-1.65}$&(1)\\
04A&4.90 (0.10)&13.34$^{+0.82}_{-0.87}$&(1)\\
04et&4.77 (0.07)&13.46$^{+0.75}_{-0.74}$&(1)\\
05cs&4.38 (0.07)&11.68$^{+1.12}_{-1.65}$&(1)\\
08bk&4.53 (0.07)&11.68$^{+1.12}_{-1.65}$&(1)\\
08cn&5.10 (0.10)&14.70$^{+0.73}_{-0.74}$&(1)\\
12A&4.57 (0.09)&13.76$^{+0.89}_{-0.83}$&(1)\\
12aw&4.92 (0.12)&15.09$^{+0.56}_{-0.57}$&(1)\\
12ec&5.16 (0.07)&15.30$^{+0.59}_{-0.59}$&(1)\\
13ej&4.69 (0.07)&15.64$^{+0.66}_{-0.65}$&(1)\\
17eaw&5.05 (0.10)&14.33$^{+1.42}_{-1.61}$&(2)(3)\\
20jfo&4.10 (0.40)&11.90$^{+1.25}_{-1.30}$&(4)\\
23ixf&5.00 (0.10)&14.99$^{+1.21}_{-1.34}$&(5)\\
\enddata
\caption{SNe II with both nebular spectroscopy and pre-SN images. References: (1)\,\citet{DB18}; (2)\,\citet{rui19_17eaw}; (3)\,\citet{2017eaw}; (4)\,\citet{kilpatrick23}; (5)\,\citet{vandyk24}.} 
\end{deluxetable}

To investigate the correlation between these two quantities, we conduct a Monte Carlo simulation: in each trial, a $M_{\rm ZAMS,neb}$ value is randomly sampled from the parent distribution (examples shown in Figure~\ref{fig:mass_distribution_examples}), and a log\,$L$ value is randomly drawn from a Gaussian distribution with the uncertainties quoted in the source papers. We then measure the Spearman's correlation coefficient $\rho$ and the significant level $p$ for this random sample. The process is repeated for 10,000 times, and we find $\rho$\,=\,0.65$^{+0.11}_{-0.12}$ and $p$\,\( < \) 0.016$^{+0.054}_{-0.013}$ with the quoted uncertainties representing the 68\% CI. 

While the correlation is reasonably strong, the significance level does not always fall below the 0.05 threshold. We identify SN 2013ej as a potential outlier. The pre-SN images suggest an $M_{\rm ZAMS}$ of 10-12\,$M_{\odot}$ based on the \texttt{KEPLER} MLR. However, SN 2013ej shows strong [O I] emission and a bright plateau phase, suggesting a highly energetic explosion. In a forthcoming work, we will show that, if SN 2013ej is indeed from a relatively low-mass progenitor, the explosion energy would need to be around 1 foe (10$^{51}$\,erg) to explain the bright plateau, which is close to the observed upper limit for SNe II (see, for example, Figure 10 of \citealt{fang24a}). \citet{nagao24} also suggest SN 2013ej belongs to a group of energetic outliers. Such high energy is not favored for low-mass progenitors in neutrino-driven explosion models (see, e.g. \citealt{stockinger20,burrows21,burrows24a,burrows24b,janka24})\footnote{We note that other mechanisms may trigger such high-energy explosions for low-mass progenitors; see the discussion in \citet{socker24}}. Removing SN 2013ej from the sample significantly improves the correlation: repeating the aforementioned MC test without SN 2013ej returns $\rho$\,=\,0.75$^{+0.10}_{-0.14}$ and $p$\,\( < \)\,0.005$^{+0.031}_{-0.004}$. However, the result is not affected by the further exclusion of any other SNe. Given these considerations, we conclude that SN 2013ej should be excluded from the sample of RSG images.

The ZAMS mass measured from nebular spectroscopy, $M_{\rm ZAMS,neb}$, is transformed to log\,$L$ through the following procedure: We conduct 10,000 Monte Carlo simulations, where in each trial, for each SNe, an $M_{\rm ZAMS,neb}$ value is randomly drawn from its $M_{\rm ZAMS, neb}$ distribution (Figure~\ref{fig:mass_distribution_examples}). For objects with detected progenitor RSG  (excluding SN 2013ej), luminosities are randomly sampled from Gaussian distributions based on the uncertainties in Table~\ref{tab:preSN}. A linear regression is then performed on the overlapping objects in these two random samples to establish the $M_{\rm ZAMS, neb}$-log\,$L$ relation in each trial, which is then applied to estimate the luminosities of the remaining SNe. 

\begin{figure}[!htb]
\epsscale{1.1}
\plotone{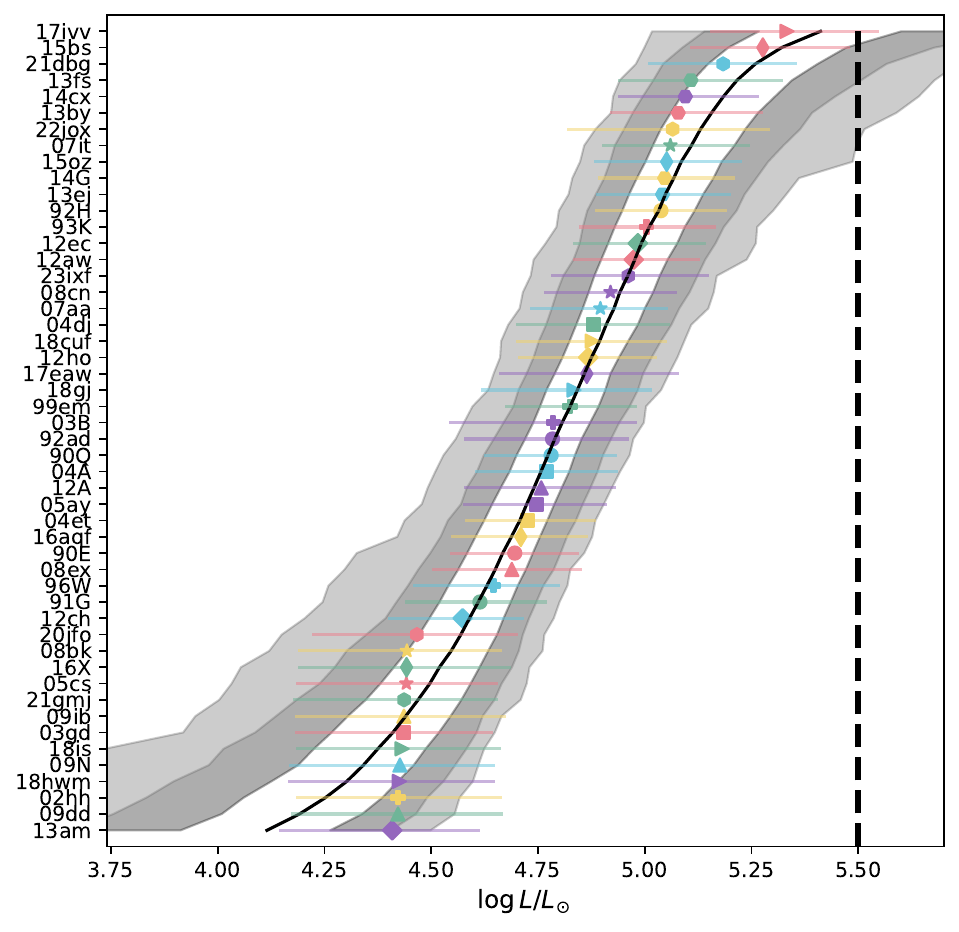}
\centering
\caption{Same as Figure~\ref{fig:neb_mass_histogram}, but for the luminosities of the progenitor RSGs inferred from the empirical MLR. The thick dashed line represents log\,$L$\,=\,5.5 dex.}
\label{fig:logL_histogram}
\end{figure}

For each individual SN II, the above Monte Carlo sampling generates 10,000 log\,$L$ values, forming a distribution that may not necessarily follow a Gaussian shape. Throughout this work, the adopted log\,$L$ is the median value of this distribution, with uncertainties defined by the 16th and 84th percentiles, which are presented in Figure~\ref{fig:logL_histogram}. No object has median log\,$L$\,\( > \)\,5.5, a value frequently quoted as the upper limit of the field RSGs (\citealt{DCB18}). The brightest progenitor is that of SN 2017ivv, with log\,$L$\,=\,5.33$^{+0.21}_{-0.18}$\,dex, indicating that the missing of bright SN II progenitor with log\,$L$\,\( > \)\,5.5\,dex is significant at 1$\sigma$ level. The luminosity distribution function (LDF) is also shown in Figure~\ref{fig:logL_histogram}, where the shaded regions represent the 68, 95 and 97.5 CI. A more detailed statistical analysis of the luminosity distribution function will be presented in \S5.

\subsection{Robustness test}
\begin{figure}[!htb]
\epsscale{1.1}
\plotone{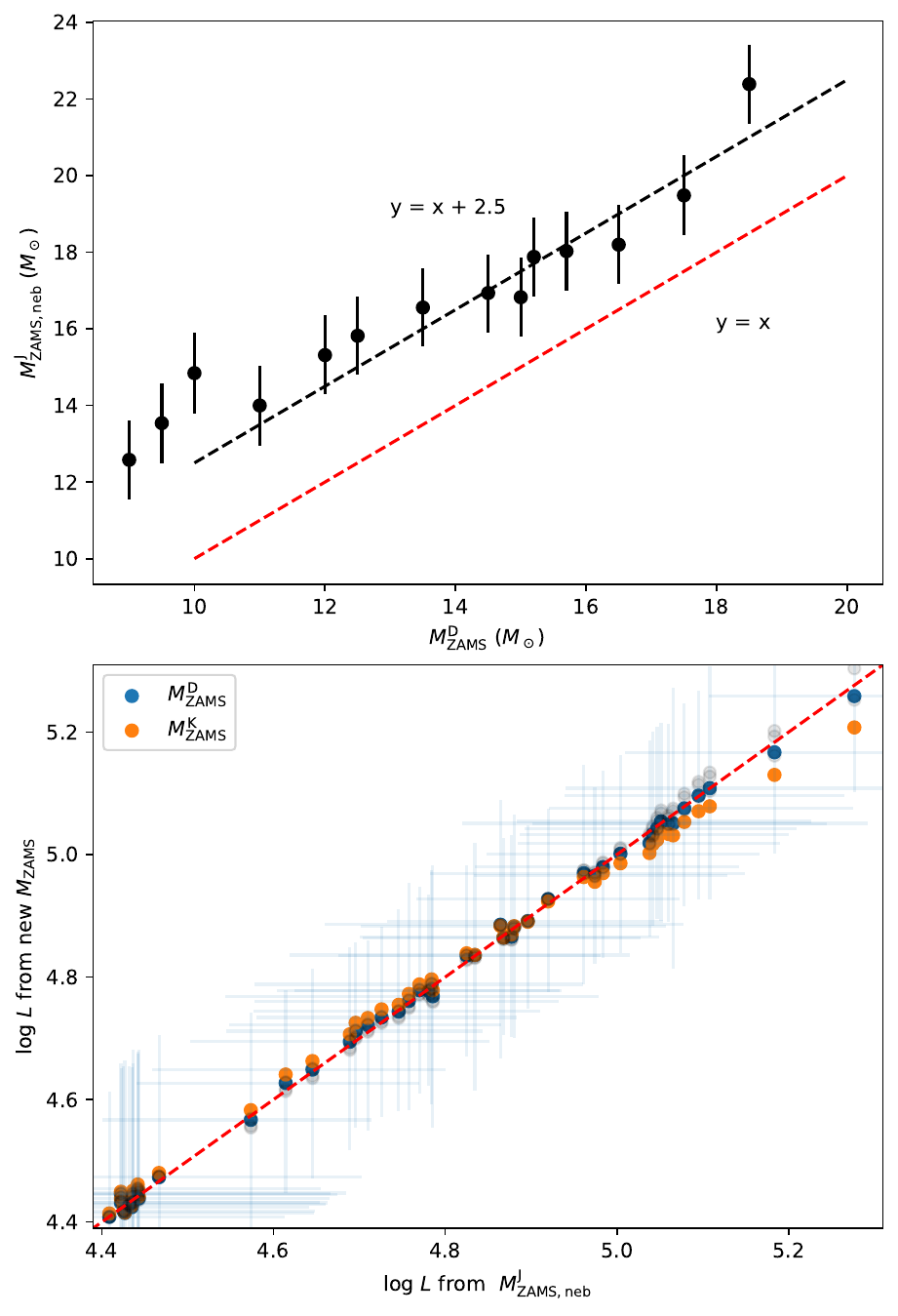}
\centering
\caption{Upper panel: Comparison between the $M_{\rm ZAMS}^{\rm D}$ of the nebular spectral models from \citet{dessart21} and their measured $M_{\rm ZAMS,neb}^{\rm J}$ using the method in the work; Lower panel: Comparison between log\,$L$ estimated from $M_{\rm ZAMS,neb}^{\rm J}$ with log\,$L$ estimated from other forms of $M_{\rm ZAMS}$. The black transparent points are measurements for several pairs of ($k\,,r$) applied in Equation~\ref{eq:general_mass}. In both panels, red dashed line represents $y$\,=\,$x$.}
\label{fig:variation_in_zams}
\end{figure}

Finally, we conduct a robustness test on the derived log\,$L$ values based on this method. In Figure~\ref{fig:logl_mco}, the 68\% CI of the empirical $M_{\rm ZAMS,neb}$-log\,$L$ relation is shown as the shaded region, while the dashed line represents the MLR of the \texttt{KEPLER} models. The observed track appears sharper than the model prediction. This discrepancy indicates a systematic offset if the uncalibrated MLR from the \texttt{KEPLER} models is directly applied to transform $M_{\rm ZAMS, neb}$ to log\,$L$. The origin of this inconsistency could arise from several uncertainties:
\begin{itemize}
    \item The nebular spectral model may depend on the details of the radiative transfer code. \citet{dessart21} introduce a set of nebular spectral models calculated by \texttt{CMFGEN}, where they employ progenitor models computed by the \texttt{KEPLER} code, similar to \citet{jerk12}, but with variations in explosion energy and a different treatment of material mixing (see also \citealt{lisakov17}). These models only cover $t$\,=\,350\,days after the explosion, so they are not compared with the observation in this work. Instead, they are treated as observed spectra, pre-processed following the procedures introduced in \S2, and the corresponding $M_{\rm ZAMS,neb}$ are measured using the same method in \S3 to test how $M_{\rm ZAMS,neb}$ is affected by the specific of the spectral models. In the upper panel of Figure~\ref{fig:variation_in_zams}, we compare the $M_{\rm ZAMS}^{\rm D}$ of the models in \citet{dessart21} with the measured $M_{\rm ZAMS,neb}^{\rm J}$ using the models in \citet{jerk12}, where we find a systematic offset, and in the range of $M_{\rm ZAMS}^{\rm D}$\,\(<\)\,20\,$M_{\rm \odot}$, the relation can be roughly expressed as
    \begin{equation}\label{eq:dessart_mass}
    M_{\rm ZAMS,neb}^{\rm J}\,=\,M_{\rm ZAMS}^{\rm D}\,+2.5,
    \end{equation}
    as shown in the upper panel of Figure~\ref{fig:variation_in_zams}.Throughout this section, all masses are given in solar mass units. The prefixes ‘D’ and ‘J’ indicate models from \citet{dessart21} and measurements based on models from \citet{jerk12,jerk14}, respectively.
    \item Uncertainties in the initial conditions. The nebular spectroscopy models employed in this work (\citealt{jerk12,jerk14}) assume an explosion energy of 1.2\,$\times$\,10$^{51}$\,erg, while most SNe II have explosion energy below this value from plateau phase light curve modeling (Figure 10 of \citealt{fang24a}). As discussed in \citet{jerk18}, estimating the mass of the emitting element from line luminosity relies on the assumption of ‘all else constant’. If progenitors with lower helium core mass (lower log\,$L$) tend to have smaller explosion energy (see, e.g., \citealt{burrows24a}), then using models with a fixed explosion energy at 1.2\,$\times$\,10$^{51}$\,erg may lead to an overestimation of $M_{\rm ZAMS,neb}$, as exemplified by the comparison of the M9 and M12 models in Figure~\ref{fig:track} (see also discussion in \citealt{jerk18}). In such a scenario, objects in the lower-left region of Figure~\ref{fig:logl_mco} would shift further left, effectively flattening the observed $M_{\rm ZAMS}$-log\,$L$ relation and making it more consistent with the MLR predicted by the \texttt{KEPLER} models. Applying the transformation 
    \begin{equation}\label{eq:kepler_mass}
    M_{\rm ZAMS}^{\rm K}\,=2.35\,\times\,M_{\rm ZAMS,neb}^{\rm J}\,-18.47,
    \end{equation}
    the observed MLR aligns with the \texttt{KEPLER} model predictions. This empirical relation accounts for all factors—such as variations in explosion energy, the $M_{\rm He\,core}$-$M_{\rm O}$ relation or material mixing—that may cause the observed MLR to deviate from \texttt{KEPLER} predictions, assuming these effects are primarily determined by the helium core properties.
\end{itemize}
\bigskip
Although $M_{\rm ZAMS,neb}$ is subject to several uncertainties as discussed above, the inferred log\,$L$ is not significantly affected. To demonstrate this, we transform $M_{\rm ZAMS,neb}$ of SNe II in the sample to new values $M_{\rm ZAMS}^{\rm D}$ with Equations~\ref{eq:dessart_mass} (which accounts for the uncertainty associated with different model sets), and to $M_{\rm ZAMS}^{\rm K}$ with Equations~\ref{eq:kepler_mass} (which accounts for the uncertainty associated with the initial condition), after which SNe II in Table~\ref{tab:preSN} are employed to establish the empirical MLR based on $M_{\rm ZAMS}^{\rm D}$ and $M_{\rm ZAMS}^{\rm K}$ respectively. The log\,$L$ values of all other objects in the sample are estimated using these updated relations based on their $M_{\rm ZAMS}^{\rm D}$ (or $M_{\rm ZAMS}^{\rm K}$) values. As shown in Figure~\ref{fig:variation_in_zams}, the newly estimated log\,$L$ are consistent with those based on $M_{\rm ZAMS,neb}^{\rm J}$ within uncertainty. More generally, we have tested the transformation
\begin{equation}\label{eq:general_mass}
    M_{\rm ZAMS}^{\rm new}\,=\,k\,\times\,M_{\rm ZAMS,neb}^{\rm J}\,+\,r
\end{equation}
for several pairs of ($k\,,r$), and we find that, as long as $k$\,\(>\)\,0 and the transformed $M_{\rm ZAMS}^{\rm new}$ values for SNe II in the sample falls within 9 to 25\,$M_{\rm \odot}$, the inferred log\,$L$ remains virtually unchanged, ensuring that the discussion in \S5 is not significantly affected by the uncertainties in the spectral or stellar evolution models.

\section{Implication for the RSG problem}
In this section, we discuss the global properties of the log\,$L$ distribution of the RSG progenitor, estimated from $M_{\rm ZAMS,neb}$ using the calibrated MLR. There are mainly two methods to assess the significance of the RSG problem, (1) Comparison of the LDF of RSG progenitors with that of RSGs in other galaxies (field RSGs). See \citet{R22} as an example; (2) Modeling the LDF of RSG progenitors using analytical functions, typically power-law forms with upper and lower cutoffs. These methods have distinct focuses as well as pros and cons. While method (1) has the advantage of being model independent, it also has several limitations: 
\begin{itemize}
\item Key focus: Using method (1), \citet{R22} concludes that the $N$(log\,$L$ \( > \)\,5.1)/$N$(log\,$L$ \( > \)\,4.6) ratio of SNe II progenitors is smaller the that of the RSGs in Large Magellanic Cloud (LMC) and other galaxies, implying fewer bright RSG than expected. Here $N$(log\,$L$ \( > \)\,4.6) refers to the number of RSGs with log\,$L$ \( > \)\,4.6 dex, and similarly for $N$(log\,$L$ \( > \)\,5.1). However, this approach does not directly address the existence of an upper luminosity cutoff in the progenitor population. The RSG problem fundamentally concerns the presence of such a cutoff, which could result from factors like explodability or pre-supernova mass-loss mechanisms (discussed later in \S6).
\item Evolutionary presentation: The field RSGs are typically less evolved than the progenitors of SNe II, which raises questions about whether they really represent the final evolutionary states of massive stars. For example, RSGs with luminosities as high as log\,$L\,\sim$\,5.5 dex may remain near the Hayashi line during helium burning and could be observed as a field RSG, but it could transition away from this state in later evolutionary phases due to processes like mass stripping from eruptive activities. As a result, they may not explode as SNe II, potentially relaxing the observed difference in the $N$(log\,$L$ \( > \)\,5.1)/$N$(log\,$L$ \( > \)\,4.6) ratio between progenitor RSGs and field RSGs.  

\end{itemize}

Given these considerations, we employ method (2) for the statistical analysis of the LDF, similar to the one applied in \citet{DB20} (sorting method). For further details, readers are encouraged to refer to that paper. Here, we briefly describe the workflow:

(1) In the previous section, we have established the LDF using a Monte Carlo method. Especially, we derive the distribution of the luminosity of the $i$-th SN, denoted as $P_{i,\rm obs}$\,(log\,$L$);

(2) Next, we construct the model LDF, which is in the power-law form
\begin{equation}
dN/d {\rm log}\,L \propto L^{1 + \Gamma_{L}},
\end{equation}
bounded by the lower limit $L_{\rm low}$ and the upper limit $L_{\rm up}$. For each set of \{$\Gamma_{L},{\rm log}\,L_{\rm low},{\rm log}\,L_{\rm up}$\}, we again estimate the LDF using the same method described in the previous section: we perform 10,000 Monte Carlo simulation, and in each trial, we draw 50 (the observed sample size in this work) log\,$L$ values from the bounded power-law distribution (representing the scatter points in Figure~\ref{fig:logL_histogram}). The uncertainties are assigned according to the ranks. For example, for the faintest progenitor in the random sample, it is assumed to follow the same log\,$L$ distribution as SN 2013am established through Monte Carlo sampling in \S4.2, but shifted by a constant to align the median value. For each simulated SN, a value is randomly drawn from its associated log\,$L$ distribution, and the full sample is sorted again. This second sorting step mimics the ranking method used in \citet{DB20}. The luminosity distribution of the $i$-th SN from the 10,000 trials is denoted as $P_{i,\rm model}$\,(log\,$L$);

\begin{figure}
\epsscale{1.1}
\plotone{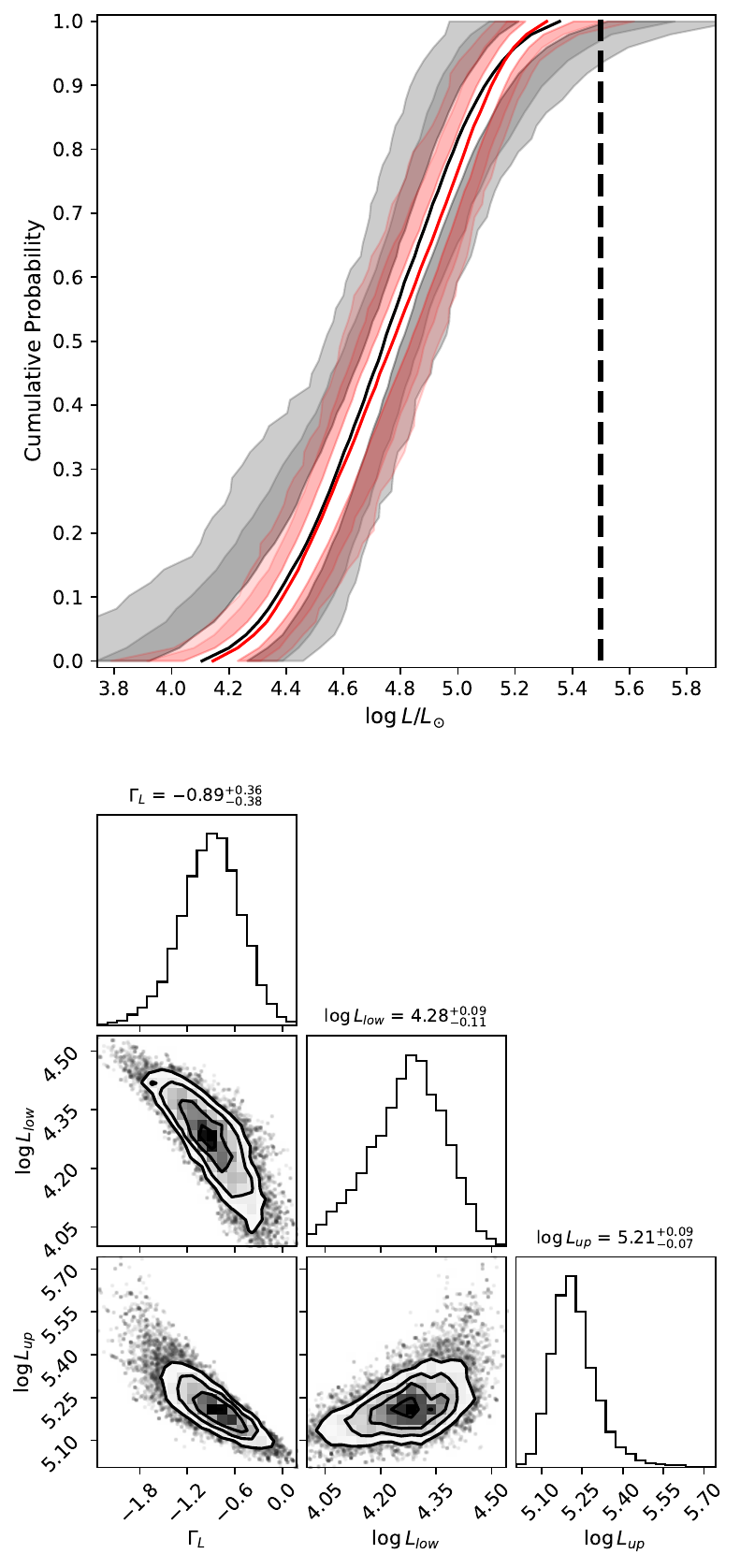}
\centering
\caption{Upper panel: The comparison of the observed LDF (black) and the model LDF with $\Gamma_{L}$\,=\,-0.89, log\,$L_{\rm low}$\,=\,4.28 and log\,$L_{\rm up}$\,=\,5.21 (red). The uncolored regions surrounding the solid lines and the transparent regions are 68, 95 and 99.7\% CIs. The thick dashed line represents log\,$L$\,=\,5.5 dex; Lower panel: the corner plot of the \texttt{emcee} routine.}
\label{fig:LDF_3para}
\end{figure}

(3) For a fixed set of \{$\Gamma_{L},{\rm log}\,L_{\rm low},{\rm log}\,L_{\rm up}$\}, the probability that the model produces the observed log\,$L$ of the $i$-th progenitor is 
\begin{equation}
P_i\,=\,\sum_{{\rm log}\,L} P_{i,\rm model}\,({\rm log}\,L)\,\times\,P_{i,\rm obs}\,({\rm log}\,L).
\end{equation}
The likelihood function is written as
\begin{equation}
{\rm ln}\,\mathcal{L} = \sum_{i} {\rm ln}\,P_{i}.
\end{equation}
After the likelihood function is established, we use the \texttt{Python} package \texttt{emcee} to infer the optimized \{$\Gamma_{L},{\rm log}\,L_{\rm low},{\rm log}\,L_{\rm up}$\} and the associated uncertainties (\citealt{emcee}). The setup is as follow: we use 60 walkers (20 walkers per parameter), and their initial positions \{$\Gamma_{L},{\rm log}\,L_{\rm low},{\rm log}\,L_{\rm up}$\}$_{\rm 0}$ are randomly sampled from \{$U{\rm (-5, 2)},U{\rm (4.0, 4.8)},U{\rm (4.8, 6.0)}$\}, the prior distributions of these parameters. Here $U$($a,b$) represents uniform distribution between $a$ and $b$. We then allow the walkers to explore the parameter space freely using \texttt{emcee.run\_mcmc} to run for 1000 chains, after which the result converges. The corner plot of the parameters and their uncertainties are shown in Figure~\ref{fig:LDF_3para}. The optimized parameters are:
\begin{align*}
&\Gamma_{L} = -0.89^{+0.36}_{-0.38}\\
&{\rm log}\,L_{\rm low} = 4.28^{+0.09}_{-0.11}\\
&{\rm log}\,L_{\rm up} = 5.21^{+0.09}_{-0.07}.\\
\end{align*}
It is interesting to see that, although a different SNe sample and different method are employed to derive the LDF of the RSG progenitor, the results match quite well with \cite{DB20}, where they find
\begin{align*}
&\Gamma_{L} = -1.12^{+0.95}_{-0.81}\\
&{\rm log}\,L_{\rm low} = 4.39^{+0.10}_{-0.16}\\
&{\rm log}\,L_{\rm up} = 5.20^{+0.17}_{-0.11}.\\
\end{align*}

Although the optimize ${\rm log}\,L_{\rm up}$ is lower than 5.5 dex, consistent with the RSG problem, the 97.5 percentile (+2$\sigma$) of the log\,$L_{\rm up}$ distribution is 5.44 dex, and the 99.8 percentile (+3$\sigma$) is 5.63 dex. This suggests that the significance of this issue is at the 2$\sigma$ to 3$\sigma$ level.

\begin{figure}
\epsscale{1.1}
\plotone{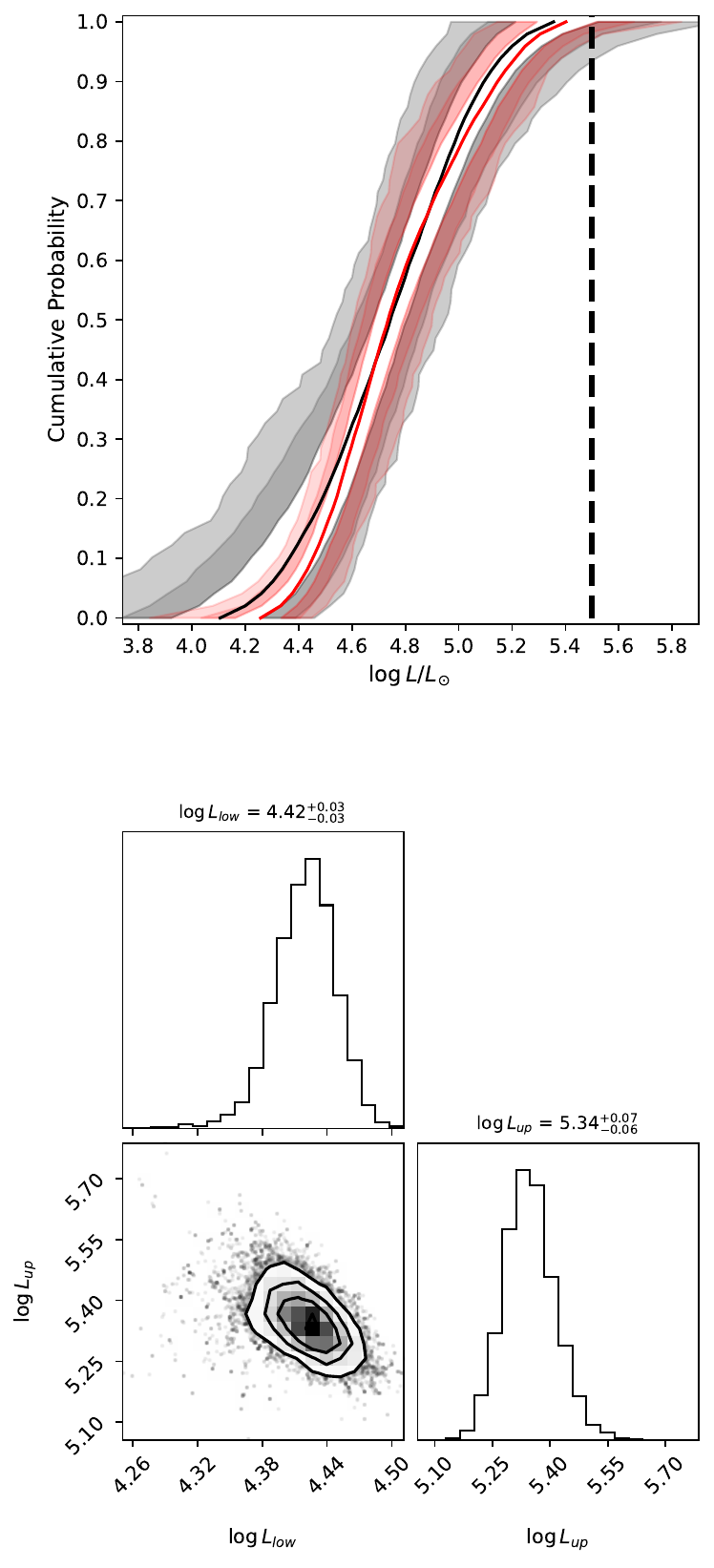}
\centering
\caption{Same as Figure~\ref{fig:LDF_3para}, but for the case when $\Gamma_{L}$ is fixed at -1.675.}
\label{fig:LDF_2para}
\end{figure}

As discussed in \citet{DB20}, part of the uncertainties in ${\rm log}\,L_{\rm up}$ and ${\rm log}\,L_{\rm low}$ arises from their degeneracy with $\Gamma_L$. Specifically, for a steeper power-law distribution, observations are more likely to detect faint objects, reducing the probability of identifying bright progenitors. This reduced detection probability allows a higher ${\rm log}\,L_{\rm up}$ to remain consistent with the data. Similarly, a sharp power-law implies a rapid increase in probability density as ${\rm log}\,L$ approaches ${\rm log}\,L_{\rm low}$. To prevent divergence at the lower end, the cutoff ${\rm log}\,L_{\rm low}$ shifts upward to maintain normalization. If the power-law index $\Gamma_{L}$ is fixed to -1.675, as would be expected if the progenitors in the sample follow the Salpeter form of initial mass function (IMF; characterized by $dN/dM_{\rm ZAMS}\,\propto\,M_{\rm ZAMS}^{-2.35}$; \citealt{imf,DB20}), the similar \texttt{emcee} routine  returns 

\begin{align*}
&{\rm log}\,L_{\rm low} = 4.42^{+0.03}_{-0.03}\\
&{\rm log}\,L_{\rm up} = 5.34^{+0.07}_{-0.06},\\
\end{align*}
as shown in Figure~\ref{fig:LDF_2para}. This result shows that fixing $\Gamma_L$ leads to higher optimized values for \{${\rm log}\,L_{\rm low},{\rm log}\,L_{\rm up}$\} compared to cases where $\Gamma_L$ is allowed to vary freely. However, the corresponding uncertainties in these parameters decrease because the degeneracy between $\Gamma_L$ and the cutoffs is removed. As a result, the RSG problem persists at a significance level of 2$\sigma$ (5.49 dex at 97.5 percentile) to 3$\sigma$ (5.57 dex at 99.8 percentile).

\begin{figure*}[!htb]
\epsscale{1.2}
\plotone{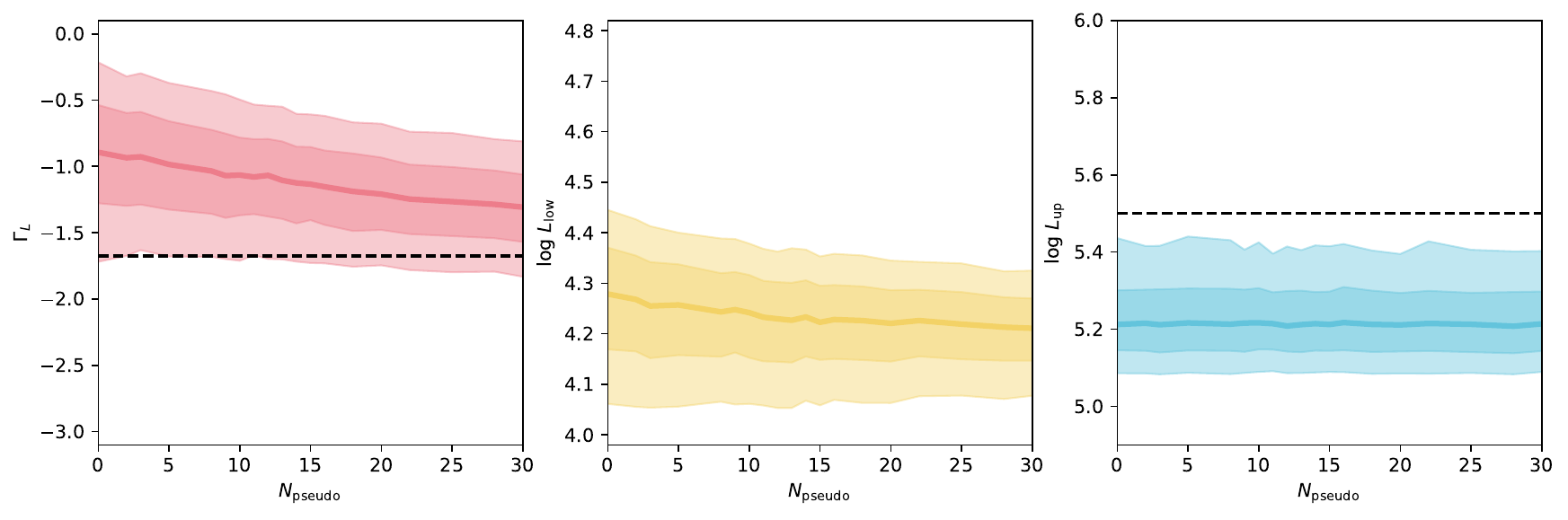}
\centering
\caption{The distributions of $\Gamma_{L}$ (left), log\,$L_{\rm low}$ (middle) and log\,$L_{\rm up}$ (right) from \texttt{emcee} routine as functions of the number of pseudo SNe, artificial data points that represents the faint RSG progenitors ($N_{\rm pseudo}$; see main text for definition). The solid lines represent the median values and the shaded regions represent 68 and 95\% CI. In the left panel, the black dashed line is $\Gamma_{L}$\,=\,-1.675; in the right panel, the black dashed line is log\,$L$\,=\,5.5 dex.}
\label{fig:pseudo_SN}
\end{figure*}
One major concern is the completeness of the sample. Indeed, if the progenitor sample adheres to the Salpeter IMF, the expected $\Gamma_{L}$ is -1.675, a value within the $1\sigma$ uncertainty reported by \citet{DB20}. In this work, the sample size is increased to $N$\,=\,50. While the uncertainties of the luminosity cutoffs are comparable to \citet{DB20}, $\Gamma_L$ is now constrained to a narrower range, and the expected value of -1.675 falls outside the 68\% CI. The relatively flat LDF suggests that the sample is probably incomplete, likely missing some low-luminosity progenitors. To address this, we account for sample incompleteness in two ways:

(1) {\bf Excluding Low-Luminosity Progenitors.} For the first attempt, we only consider bright progenitors. Previous studies suggest that for progenitors with log\,$L \gtrsim 4.6$–4.7, the LDF is consistent with those of field RSGs \citep{R22,strotjohann24,healy24}\footnote{It should be cautious that, field RSGs are less evolved than SN progenitors, raising questions about whether they truly reflect the properties of RSGs at the onset of core collapse.}. Further, for SN progenitors with log\,$L$ \( < \) 4.6, the measured luminosity may be unreliable, otherwise some SNe would have $M_{\rm ZAMS}$ \( < \) 8\,$M_{\rm \odot}$, below the minimum mass required for a single star to undergo core collapse \citep{heger03}. Based on these considerations, we adopt a cutoff at log\,$L = 4.6$, retaining only bright SNe progenitors. This reduces the sample size to $N$\,=\,33, comparable to the sample size of \citet{DB20}. Repeating the \texttt{emcee} routine gives  
\begin{align*}
&\Gamma_{L} = -1.67^{+1.50}_{-1.14}\\
&{\rm log}\,L_{\rm low} = 4.67^{+0.05}_{-0.10}\\
&{\rm log}\,L_{\rm up} = 5.25^{+0.26}_{-0.12}.\\
\end{align*}
The smaller sample size relaxes parameter constraints, increasing uncertainties to levels comparable to \citet{DB20}. Notably, the expected value of -1.675 now falls within the uncertainty of $\Gamma_L$, while the optimized value of ${\rm log}\,L_{\rm up}$ increases by 0.04 dex (10\% in linear scale). Consequently, the significance of the RSG problem is reduced to below 1$\sigma$;

(2) {\bf Including pseudo SNe for low-luminosity progenitors.} In the second approach, we address the faint side of the LDF. While these low-luminosity progenitors do not directly affect the estimate of log\,$L_{\rm up}$, they play a crucial role in constraining $\Gamma_{L}$ and log\,$L_{\rm low}$. By changing the overall shape of the LDF, they can indirectly affect the estimate of log\,$L_{\rm up}$. For instance, decreasing $\Gamma_L$ can result in higher values of ${\rm log}\,L_{\rm up}$ as discussed above.

To investigate this effect, we introduce pseudo SNe, artificial data points representing the missing low-luminosity progenitors. These pseudo SNe are assigned the same $M_{\rm ZAMS,neb}$ distribution as objects below the M12 track (middle panel of Figure~\ref{fig:mass_distribution_examples}). The exact luminosity distribution of these missing progenitors is unknown, and our assignment is somewhat arbitrary. However, it is reasonable: as shown in Figure~\ref{fig:logl_mco}, the faintest progenitors correspond to SNe located below the M12 track, most of which are low-luminosity SNe II. These SNe are characterized by low $^{56}$Ni production, making them faint and difficult to detect during the nebular phase \citep{pastorello04,spiro14,murai24}, and thus our nebular spectra sample in this range is probably incomplete. By varying the number of pseudo SNe ($N_{\rm pseudo}$) added to the observed sample, we repeat the previous analysis to infer the optimized parameters \{$\Gamma_{L},{\rm log}\,L_{\rm low},{\rm log}\,L_{\rm up}$\} with the \texttt{emcee} routine. The results, along with their 68\% and 95\% CIs, are shown in Figure~\ref{fig:pseudo_SN}.

As expected, $\Gamma_L$ and log\,$L_{\rm low}$ decrease with the increase of $N_{\rm pseudo}$, while ${\rm log}\,L_{\rm up}$ remains largely unaffected. For $N_{\rm pseudo}$\,=\,30, when the number of the pseudo SNe becomes comparable to the observed sample, we obtain 
\begin{align*}
&\Gamma_{L} = -1.31^{+0.25}_{-0.26}\\
&{\rm log}\,L_{\rm low} = 4.21^{+0.07}_{-0.06}\\
&{\rm log}\,L_{\rm up} = 5.21^{+0.09}_{-0.07}.\\
\end{align*}
At first glance, the result might seem contradictory to the earlier discussion, where fixing $\Gamma_L$ to -1.675 resulted in larger log\,$L_{\rm up}$. However, the key distinction lies in sample size. Adding pseudo SNe effectively enlarges the sample. While lower $\Gamma_L$ (a sharper power-law distribution) biases detections toward low-luminosity progenitors, allowing for larger ${\rm log}\,L_{\rm up}$, a larger sample size reduces the likelihood of missing high-luminosity progenitors. The competition between these factors stabilizes ${\rm log}\,L_{\rm up}$ at an approximately constant value. Thus, the RSG problem persists at a significance level of $2\sigma$ to $3\sigma$, as shown in the right panel of Figure~\ref{fig:pseudo_SN}.

\section{Discussion}
In \citet{morozova18} and \citet{martinez22}, the cutoff masses of the $M_{\rm ZAMS}$ mass distribution for large samples of SNe II were investigated through plateau-phase light curve modeling. Converting the luminosity cutoffs derived in this work into $M_{\rm ZAMS}$ values is essential for a direct comparison with these studies and for assessing the robustness of the results. We perform this conversion using the MLR from the \texttt{KEPLER} models, i.e., first convert the luminosity scales (say, log\,$L_{\rm up}$) to $M_{\rm He\,core}$ via Equation~\ref{eq:Mhe_L}, which is subsequently converted to $M_{\rm ZAMS}$ using the $M_{\rm ZAMS}$-$M_{\rm He\,core}$ relation in \citet{sukhbold16}. Although this introduces uncertainties associated with different stellar evolution codes (as discussed in \S4), the progenitor models of \citet{morozova18} and \citet{martinez22} follow a similar MLR, allowing for a fair comparison.

\begin{figure}[!htb]
\epsscale{1.2}
\plotone{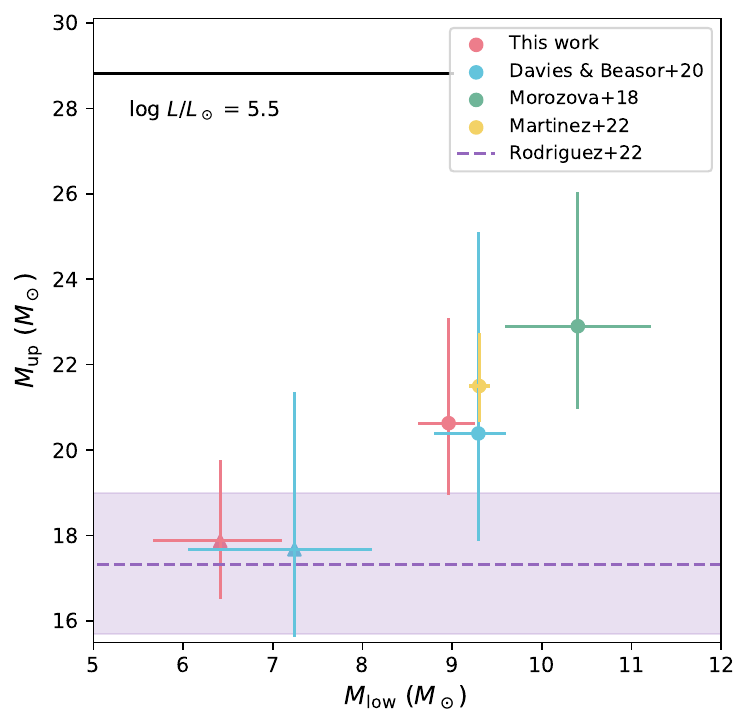}
\centering
\caption{The comparison of upper and lower cutoffs of $M_{\rm ZAMS}$ inferred using different methods. The pink and light blue circles are $M_{\rm ZAMS}$ converted from the luminosities cutoffs in this work and \citet{DB20}, using the MLR of \texttt{KEPLER} models, while the triangles are converted using the MLR described in \citet{DB18}. The purple dashed line and the transparent region represent $M_{\rm ZAMS}$ converted from log\,$L$\,=\,5.063\,$\pm$\,0.077 using the MLR of \texttt{KEPLER} models, corresponding to the maximum log\,$L_{\rm prog}$($L_{\rm [O I]}$) in \citet{R22}. The black solid line corresponds to log\,$L$\,=\,5.5, the empirical upper luminosity of field RSG.}
\label{fig:compare_mass}
\end{figure}

The upper mass cutoff ($M_{\rm up}$) converted from log\,$L_{\rm up}$ derived in this work is 20.63$^{+2.42}_{-1.64}$\,$M_{\rm \odot}$, where the quoted uncertainties define the 68\% CI. Similarly, we find a lower mass cutoff at $M_{\rm low}$\,=\,8.95$^{+0.27}_{-0.32}$\,$M_{\rm \odot}$.\footnote{For log\,$L\,<\,$4.3 (corresponding to $M_{\rm ZAMS}$\,=\,9\,$M_{\rm \odot}$), the estimation is based on extrapolation.} In Figure~\ref{fig:compare_mass}, we compare the $M_{\rm up}$ and $M_{\rm low}$ with the measurements from plateau phase light curve modeling (\citealt{morozova18,martinez22}). The purple strip is the $M_{\rm ZAMS}$ converted from the maximum log\,$L_{\rm prog}$($L_{\rm [O I]}$) in \citet{R22} (5.063\,$\pm$\,0.077; see Table 9 in that paper). 

Despite the use of different methods, including SN progenitor luminosity analysis (\citealt{DB18, DB20}), plateau-phase light curve modeling (\citealt{morozova18, martinez22}), and nebular-phase spectroscopy (\citealt{R22}; this work), the inferred $M_{\rm up}$ consistently falls within the range of $\sim$\,18 to 23\,$M_{\rm \odot}$, well below $M_{\rm ZAMS}$\,$\sim$\,29.4\,$M_{\rm \odot}$ converted from log\,$L$\,=\,5.5. Although the discrepancy between $M_{\rm up}$ and the threshold 29.4\,$M_{\rm \odot}$ is at the level of 1 to 3$\sigma$, its persistence across different methodologies suggests that it may reflect a real physical problem. It would be interesting to explore the physical implications of this discrepancy.

Several theories have been proposed to explain the dearth of SNe II with luminous progenitor. The first one involves the "explodability" of massive stars. \citet{sukhbold18} studied the core structures of a large grid of progenitors with varying physical parameters and found that the upper mass limit for SNe II appears to converge at $\sim$\,20\,$M_{\rm \odot}$. In their models, progenitors with $M_{\rm ZAMS}$\,$\sim$\,20 to 25\,$M_{\rm \odot}$ experience collapse into black holes, resulting in failed SNe (see also Table 6 and Figure 19 of \citealt{sukhbold16}). However, multi-dimensional models from \citet{burrows24a} suggest successful explosions in this mass range. In a subsequent study, \citet{burrows24b} showed that even for a 23\,$M_{\rm \odot}$ progenitor that forms a black hole, the model produces an explosion rather than remaining quiescent. Thus, the role of explodability in explaining the absence of massive progenitors for SNe II remains a topic of ongoing debate.

Another hypothesis involves the surface properties of RSG progenitors. Under this scenario, massive stars may still explode but as SNe types other than SNe II. According to the RSG models in \citet{sukhbold16}, single-star evolution predicts that the hydrogen-rich envelope is fully removed by stellar winds only in stars with $M_{\rm ZAMS}$\,$\gtrapprox$\,30\,$M_{\rm \odot}$. However, if mass-loss from stellar winds is stronger than assumed in current stellar evolution models, this threshold could be reduced to $\sim$\,20\,$M_{\rm \odot}$ (\citealt{meynet15}). There is also observational evidence suggesting that more massive stars are more likely to form in close binary systems (\citealt{moe17, moe19}). Such systems can efficiently strip the hydrogen-rich envelope, skewing the mass (and luminosity) distribution of SNe II progenitors toward lower values. Further, a luminosity of log\,$L\,\sim$\,5.2 dex (log\,$L_{\rm up}$ found in this work and \citealt{DB20}) is already sufficient to trigger pulsation (\citealt{soraisam18,goldberg20}). While pulsation-driven mass-loss is not included in stellar evolution codes, \citet{yoon10_pulse} demonstrated that once initiated, it becomes a runaway process capable of stripping nearly the entire hydrogen envelope. In this scenario, the pulsating RSGs would eventually explode as stripped-envelope SNe or interacting SNe (\citealt{smith09}) rather than SNe II. In a recent work, \citet{cheng24} propose an eruptive mass-loss mechanism, under which progenitor models with $M_{\rm ZAMS}$\,$\gtrapprox$\,20\,$M_{\rm \odot}$ will end their lives as blue supergiant. This could explain the apparent absence of RSG progenitors above this mass range. Further investigations on the instabilities of massive stars are important to fully understand these processes and their potential connections with the RSG problem.

\section{Conclusion}
The RSG problem—namely, the observed absence of luminous RSG progenitors for SNe II—raises fundamental challenges about our understanding of massive star evolution and SN explosion mechanisms. In this work, we investigate this topic through a statistical analysis of nebular spectroscopy for a large sample of SNe II. Since nebular spectroscopy provides an independent estimate of the $M_{\rm ZAMS}$, it offers an important cross-check on the RSG problem, complementing results from pre-SN imaging of RSG progenitors.

To achieve this, we first estimate $M_{\rm ZAMS,neb}$ for individual SNe by comparing the fractional flux of the [O I] emission emerging in the nebular spectroscopy with spectral models. The resulting $M_{\rm ZAMS,neb}$ values are then compared with the observed luminosities of RSG progenitors for SNe with detected progenitors, revealing a strong and statistically significant correlation. Using this empirically calibrated mass-luminosity relation, we establish the progenitor luminosity distribution function (LDF) for the full sample. The brightest progenitor in our sample is that of SN 2017ivv, with log\,$L$\,=\,5.33$^{+0.21}_{-0.18}$\,dex. Notably, there is no progenitor exceeding log\,$L$\,\( >\)\,5.5\,dex—the empirical upper luminosity limit for field RSGs—at a significance level of approximately 1$\sigma$.

The LDF is modeled using a power-law function with upper and lower luminosity cutoffs, adopting a Monte Carlo method similar to that of \citet{DB20}. Despite differences in sample selection and methods for measuring log\,$L$, our results are consistent with \citet{DB20}. We find an upper luminosity cutoff of log\,$L_{\rm up}$\,=\,5.21$^{+0.09}_{-0.07}$\,dex, with the absence of progenitors above log\,$L$\,=\,5.5\,dex being statistically significant at the 2$\sigma$ to 3$\sigma$ level.

Finally, we convert the luminosity cutoffs derived in this work back to the $M_{\rm ZAMS}$ scale using the mass-luminosity relation from \texttt{KEPLER} models and compare these with constraints from plateau light curve modeling. Despite methodological differences, both approaches consistently indicate an upper mass cutoff for SNe II progenitors below 29\,$M_{\rm \odot}$ at a significance level of 1–3$\sigma$. While each individual method provides only marginal significance, their consistency suggests that the lack of luminous RSG progenitors is likely a real physical problem. This finding highlights the need for further investigations on the explodability of high-mass progenitors and the late-phase instabilities of massive stars to fully understand their potential connections to the RSG problem.

\begin{acknowledgements}
The authors thank the referee for comments that helped improve the manuscript.
The authors are grateful to Yize Dong and Masaomi Tanaka for providing the nebular spectroscopy of SNe 2018cuf and 2021gmj, and to Koh Takahashi for sharing the \texttt{HOSHI} models. 

Q.F. acknowledges support from the JSPS KAKENHI grant 24KF0080. T.J.M is supported by the Grants-in-Aid for Scientific Research of the Japan Society for the Promotion of Science (JP20H00174, JP21K13966, JP21H04997). K.M. acknowledges support from the JSPS KAKENHI grant JP20H00174, JP24H01810 and 24KK0070. SNe data used in this work are retrieved from the Open Supernova Catalog (\citealt{open_SN}), the Weizmann Interactive Supernova Data Repository (WISeREP; \citealt{wiserep}) and the Supernova Database of UC Berkeley (SNDB; \citealt{sndb_origin,SNDB}).

\end{acknowledgements}

\clearpage
\newpage
\appendix
Table~\ref{tab:SNe sample} shows the list of the SNe and the nebular spectroscopy used in this work.
\setcounter{table}{0}
\renewcommand{\thetable}{A\arabic{table}}

\startlongtable
\centerwidetable
\begin{deluxetable*}{cccccccc}
\label{tab:SNe sample}
\tablecaption{SNe II sample in this work.}
\tablehead{\colhead{Name}&$cz$&$t_{\rm exp}$&$t_{\rm obs}$&Phase&Ref.\\
\colhead{}&$\rm km\,s^{-1}$&(MJD)&(MJD)&(days)&}
\startdata
1990E&1242&47938&48268&330&(1)\\
1990Q&1905&48042&48362&320&(1)\\
1991G&757&48280&48636&356&(2)(3)\\
1992H&1794&48656&49047&391&(2)(4)\\
1992ad&1280&48805&49091&286&(2)(5)\\
1993K&2724&49074&49359&285&(2)\\
1996W&1740&50180&50478&298&(6)\\
1999em&718&51476&51793&317&(7)(8)\\
2002hh&48&52576&52972&396&(2)(9)\\
2003B&1107&52622&52897&275&(10)(11)\\
2003gd&658&52716&52940&225$^{*}$&(8)(12)\\
2004A&852&53010&53296&286&(2)(13)(14)\\
2004dj&131&53181&53442&261&(2)(15)\\
2004et&48&53270&53624&354&(8)(16)(17)\\
2005ay&810&53456&53741&285&(8)(18)(19)\\
2005cs&600&53548&53882&334&(8)(20)(21)\\
2007aa&1466&54131&54526&395&(22)(23)\\
2007it&1196&54348&54616&268&(11)(24)\\
2008bk&230&54550&54810&260&(11)(25)\\
2008cn&2592&54598&54952&354&(23)(26)\\
2008ex&3945&54694&54979&285&(2)(27)\\
2009N&1034&54848&55260&412&(28)\\
2009dd&1025&54925&55334&359&(6)\\
2009ib&1305&55041&55303&262&(29)\\
2012A&750&55933&56340&407&(30)\\
2012aw&779&56002&56334&332&(31)(32)\\
2012ch&2590&56045&56402&357&(2)(33)\\
2012ec&1408&56144&56545&401&(34)\\
2012ho&2971&56255&56573&318&(2)(35)\\
2013am&1145&56374&56653&279&(2)(36)\\
2013by&1145&56404&56691&287&(37)(38)\\
2013ej&658&56497&56834&337&(2)(39)(40)\\
2013fs&3556&56572&56840&268&(41)\\
2014G&1170&56670&57011&341&(42)\\
2014cx&1646&56902&57230&328&(43)\\
2015bs&8100&56920&57341&421&(44)\\
ASASSN15oz&2100&57262&57603&341&(45)\\
2016X&1320&57406&57746&340&(46)(47)\\
2016aqf&883&57440&57770&330&(48)\\
2017eaw&40&57886&58131&245&(49)\\
2017ivv&1680&58092&58424&332&(50)\\
2018is&1734&58133&58519&386&(51)\\
2018cuf&3270&58293&58628&335&(52)\\
2018hwm&2685&58425&58814&389&(53)\\
2020jfo&1506&58974&59282&308&(54)(55)(56)(57)\\
2021dbg&6000&59258&59611&353&(58)\\
2021gmj&994&59292&59678&386&(59)(60)\\
2022jox&2667&59707&59947&240&(61)\\
2023ixf&241&60083&60341&258&(62)(63)\\
\hline
\enddata
\tablecomments{The columns are (from left to right): SN name, recession velocity of the host galaxy, date of explosion, observed date of the nebular spectrum, phases of the nebular spectrum and references: (1)\cite{gomez00} (2)\cite{SNDB}; (3)\cite{1991G}; (4)\cite{1992H}; (5)\cite{1992ad}; (6)\cite{inserra13}; (7)\cite{1999em}; (8)\cite{faran14}; (9)\cite{2002hh}; (10)\cite{anderson14}; (11)\cite{gutierrez17}; (12)\cite{2003gd}; (13)\cite{2004A1}; (14)\cite{2004A2}; (15)\cite{2004dj}; (16)\cite{2004et1}; (17)\cite{2004et2}; (18)\cite{2005ay1}; (19)\cite{2005ay2}; (20)\cite{2005cs1}; (21)\cite{2005cs2}; (22)\cite{2007aa}; (23)\cite{maguire12}; (24)\cite{2007it}; (25)\cite{2008bk}; (26)\cite{2008cn}; (27)\cite{2008ex}; (28)\cite{2009N}; (29)\cite{2009ib}; (30)\cite{2012A}; (31)\cite{2012aw}; (32)\cite{jerk14}; (33)\cite{2012ch}; (34)\cite{jerk15}; (35)\cite{2012ho}; (36)\cite{2013am}; (37)\cite{2013by}; (38)\cite{black17}; (39)\cite{2013ej}; (40)\citet{2013ej2}; (41)\cite{2013fs}; (42)\cite{2014G}; (43)\cite{2014cx}; (44)\cite{2015bs}; (45)\cite{15oz}; (46)\cite{2016X1}; (47)\cite{2016X2}; (48)\cite{2016aqf}; (49)\cite{2017eaw}; (50)\cite{2017ivv}; (51)\cite{2018is}; (52)\cite{2018cuf}; (53)\cite{2018hwm}; (54)\cite{2020jfo}; (55)\cite{2020jfo2}; (56)\cite{2020jfo3}; (57)\cite{kilpatrick23}; (58)\cite{zhao24}; (59)\cite{murai24}; (60)\cite{2021gmj}; (61)\cite{2022jox}; (62)\cite{2023ixf}; (63)\cite{2023ixf_nebular}.}
\end{deluxetable*}

{}

\end{document}